\documentclass[
  journal=custom,
  manuscript=research-article, 
  year=2025,
  volume=00,
]{scntfcjrnl-journal}

\usepackage{microtype,siunitx,booktabs,pdflscape,rotating,array,ragged2e,multirow}
\usepackage[table]{xcolor}
\usepackage[hidelinks]{hyperref}

\usepackage{booktabs}
\usepackage{placeins}
\usepackage{threeparttable}
\usepackage{longtable}

\title[Modelling intermittent mosquito infectiousness]{A modelling perspective on mosquito infectiousness: time-varying transmission competence in arbovirus vectors}

\author{Léa Loisel}
\affiliation{Oniris, INRAE, BIOEPAR, Nantes, 44300, France}
\email[L. Loisel]{lea.loisel@inrae.fr}

\author{Tristan Monrocq}
\affiliation{Oniris, INRAE, BIOEPAR, Nantes, 44300, France}

\author{Vincent Raquin}
\affiliation{IVPC UMR754, EPHE, Université PSL, INRAE, Universite Claude Bernard Lyon1, F-69007, Lyon, France}

\author{Pauline Ezanno}
\affiliation{Oniris, INRAE, BIOEPAR, Nantes, 44300, France}

\author{Gaël Beaunée}
\affiliation{Oniris, INRAE, BIOEPAR, Nantes, 44300, France}


\begin{document}

\begin{abstract}
Mosquito vector competence is usually represented as a process in which, once virus is detected in saliva, mosquitoes are assumed to remain infectious for life, implying an irreversible transition to the transmitting state. However, a subset of vector competence experiments report declines in the proportion of transmitting mosquitoes at late times post-exposition, suggesting that transmission capacity may not be permanently maintained. To investigate this hypothesis, we extended a previously developed stochastic intra-vector viral dynamics model by introducing additional transmission states allowing either permanent cessation or temporary interruption of transmission. We fitted three competing models to longitudinal experimental data from 52 vector competence conditions covering chikungunya, dengue, Zika, West Nile, and Rift Valley fever viruses, and compared their ability to reproduce observed infection, dissemination, and transmission dynamics using Approximate Bayesian Computation with Sequential Monte Carlo inference. Among the 10 experimental conditions showing a decline in transmitter proportions over time, models allowing mosquitoes to leave the transmitting state provided a better fit in 7 cases, with clear improvement in 5 cases. In these conditions, allowing interruption of transmission also increased posterior estimates of the proportion of mosquitoes that successfully crossed all intra-mosquito barriers, whereas estimates of infected- and disseminated-state durations were largely unchanged. Estimated transmitting-state durations varied across experiments. In the few cases where intermittent transmission provided the best fit, its performance was not clearly better than that of the permanent cessation model, and the estimated non-transmitting period typically lasted several days, making the two models effectively similar over the time scale of the experiments. These results indicate that the classical assumption of lifelong mosquito infectiousness does not always provide the best explanation for vector competence data and may lead to underestimation of the proportion of mosquitoes that truly become capable of transmission. Incorporating time-varying transmission competence into intra-vector models could improve interpretation of vector competence experiments and refine epidemiological representations of arbovirus transmission.

\end{abstract}

\section{INTRODUCTION }
\label{sec:int}

Arboviruses transmitted by mosquitoes represent a major and growing threat to human and animal health worldwide \citep{gubler_global_2002, jones_global_2008,wu_arbovirus_2019}. The efficiency of their transmission depends on the vector competence of mosquitoes, a multifaceted trait that encompasses the ability of a vector to acquire a virus, support its replication and dissemination within the body, and ultimately transmit it to a susceptible host through saliva during blood feeding \citep{kramer_dynamics_2003}.

Vector competence is commonly reduced to a binary outcome in experimental studies: once a mosquito is detected with virus in its saliva, it is classified as infectious and assumed to remain capable of transmission until death. This representation implies that, after crossing the transmission barrier, the virus establishes persistent infection within the mosquito in a stable and irreversible manner. Biologically, this assumption is consistent with the fact that the mosquito innate immune response is considered to control viral replication without completely clearing the infection \citep{lee_mosquito_2019}.The mosquito immune system and its microbiota contribute to a balance between resistance and tolerance, helping to control viral load while limiting the deleterious effects of infection on mosquito fitness \citep{lambrechts_manipulating_2019,oliveira_how_2020}. In transmission models, this biological persistence is formalized as an absorbing infectious compartment, thereby constraining infectiousness to be monotonic and irreversible over time \citep{reiner_systematic_2013}.

Recent experimental evidence, however, challenges this classical paradigm. Two studies on chikungunya virus have reported a decline in the proportion of mosquitoes with detectable virus in their saliva at later times post-infection \citep{prudhomme_native_2019,robison_comparison_2020}. Similar patterns have been observed, although not discussed in detail, in studies involving other arboviruses and mosquito species \citep{amraoui_potential_2019, seixas_potential_2018,merwaiss_chikungunya_2021}. These observations suggest that the capacity of mosquitoes to transmit viruses may not be permanently maintained once acquired \citep{robison_comparison_2020}, calling into question a core assumption underlying current representations of vector competence.

The interpretation of these findings remains uncertain. One possibility is that the apparent decline in salivary infection reflects methodological artifacts rather than true biological processes. Viral loads in saliva may fluctuate around the detection threshold, leading to false-negative results, or mosquitoes may fail to expectorate saliva during sampling despite being infectious \citep{prudhomme_native_2019}. Alternatively, biological mechanisms may be involved, such as viral clearance mediated by the mosquito antiviral response or the natural death of virions within the salivary glands \citep{prudhomme_native_2019}. In addition, differential mortality between infected and uninfected mosquitoes \citep{reiskind_exposure_2010, da_silveira_zika_2018} could result in a selective loss of infectious individuals over time, thereby altering observed transmission patterns at the population level.

Beyond a permanent loss of transmission capacity, an intermediate hypothesis has gained attention: that viral excretion in saliva may be intermittent. Longitudinal experimental studies, in which individual mosquitoes are repeatedly sampled over time, have shown marked temporal variability in the quantity of virus excreted in saliva, both within the same individual and across individuals \citep{mayton_age-structured_2020}. These results support the idea that mosquito infectiousness may fluctuate over time rather than follow a simple monotonic trajectory toward permanent transmissibility.

Considering transmission competence as a dynamic, time-dependent process has consequences for our understanding of vector–host transmission dynamics. In particular, it challenges the common assumption that the extrinsic incubation period marks a one-way transition to a permanently infectious state. If mosquitoes can temporarily or permanently lose their ability to transmit, or transmit intermittently, then the effective duration of infectiousness may be substantially shorter or more heterogeneous than currently assumed. This could affect estimates of transmission potential, epidemic timing, and the impact of vector control interventions derived from epidemiological models.

Recent modelling efforts have highlighted the importance of explicitly representing intra-vector infection dynamics (IVD) when describing arbovirus transmission \citep{loisel_intra-vector_2025, fontaine_epidemiological_2018,lequime_modeling_2020}. However, IVD modelling frameworks generally assume that once the transmission barrier is crossed and the virus reaches the saliva, mosquitoes remain indefinitely infectious. This assumption is consistent with the prevailing consensus that arboviruses are not eliminated once systemic infection is established in the mosquito. Nevertheless, discrepancies between model predictions and observed data at late time points post-exposition, particularly decreases in the proportion of transmitting mosquitoes, suggest that this final stage of the IVD may itself be dynamic \citep{loisel_intra-vector_2025}. As such, the transmitter state may not be an absorbing state, but rather one from which mosquitoes could temporarily or permanently exit, or within which infectiousness may fluctuate over time.

Building on this perspective, extending IVD frameworks to allow for loss or intermittence of transmission competence represents a natural next step. Incorporating dynamic transmission states into intra-vector models provides a mean to formally test alternative biological hypotheses regarding salivary infection, while maintaining a mechanistic link between within-mosquito processes and population-level transmission. Such an extension also offers a coherent way to reassess how infectious mosquitoes are represented in vector–host models, moving beyond static classifications toward a time-resolved description of mosquito infectiousness.

In this context, the aim of this study is to explicitly investigate the hypothesis of a cessation or intermittence of mosquito transmission capacity over time. Using longitudinal experimental data and a modelling framework that captures dynamic transitions of mosquitoes between transmission states, we seek to assess whether a time-varying representation of vector competence provides a better description of observed data than the classical assumption of lifelong infectiousness. More broadly, this work contributes to ongoing efforts to integrate intra-vector biological processes into transmission models and to refine the conceptual foundations of vector competence in arbovirus epidemiology.

\section{MATERIALS AND METHODS}
\label{sec:obs}
\subsection{Experimental data}
Vector competence experimental data previously compiled for chikungunya virus (CHIKV), dengue virus (DENV), and Zika virus (ZIKV) in a previous study \citep{loisel_intra-vector_2025} were used to perform model inference. The database used was assembled through an extensive literature search and selection process described in detail in the previous study. For the present study, only experiments reporting all three intra-vector viral dynamic (IVD) stages: infected ($I$), disseminated ($D$), and transmitter ($T$), were retained. In addition, experimental data on West Nile virus (WNV) and Rift Valley fever virus (RVFV), extracted from the same initial literature database using similar inclusion criteria, were incorporated into the study. Inclusion criteria were slightly relaxed compared to those applied to CHIKV, DENV, and ZIKV; indeed, no minimum number of mosquitoes per time point was imposed, allowing the inclusion of studies with smaller sample sizes. All other selection criteria were identical to those described in \cite{loisel_intra-vector_2025}. Altogether, the selected articles resulted in a total of 22 articles encompassing 52 experimental conditions (Tables \ref{chikv_table}, \ref{denv_table}, \ref{rvfv_table}, \ref{wnv_table}, \ref{zikv_table} in Supplementary Material). Each experimental condition corresponds to a vector competence experiment conducted on female mosquitoes of a given genus, species, and geographical origin, exposed to a specific viral isolate at a single infectious dose during the blood meal. Across studies, experimental protocols were broadly comparable. Fully engorged females were maintained under controlled temperature, humidity, and light conditions, and groups of 16–185 (median = 39) individuals were sampled at specific days post-exposure (DPE). Virus detection was performed using RT-PCR, focus-forming assays (FFA), or plaque-forming assays (PFA). The presence of the virus in different mosquito tissues (e.g., thorax, abdomen, legs, wings, head, saliva) was used to classify individuals into infected, disseminated, or transmitter states at each DPE.

\subsection{Intra-vector viral dynamics (IVD) model}

\textbf{Model structure and main characteristics}

\begin{figure*}[hbt!]
\centering
\includegraphics[width=0.75\textwidth]{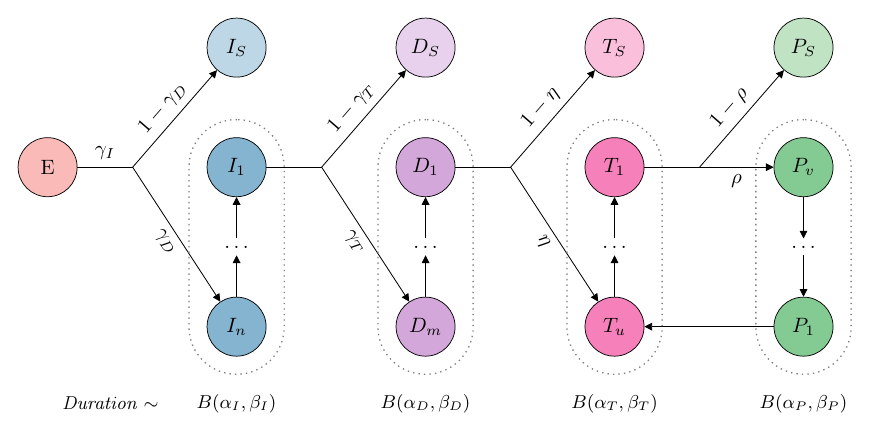}
\caption{\textbf{Conceptual diagram of the intra-vector infection dynamic model}
Each compartment represents a vector state: $E$, exposed; $I$, infected; $D$, disseminated; $T$, infectious (i.e.\ transmitting); and $P$, transmission interruption. For each state $X \in \{I,D,T,P\}$, $X_{\text{S}}$ denotes vectors remaining in state $X$, whereas $X_1$ to $X_k$ denote vectors remaining in that state for 1 to $k$ time steps (with $k=n$ for $I$, $m$ for $D$, $u$ for $T$, and $v$ for $P$). In particular, $P_{\text{S}}$ denotes vectors with permanent interruption of transmission, whereas $P_1$ to $P_v$ correspond to temporary interruption of transmission. The model parameters are $\gamma_I$, $\gamma_D$, and $\gamma_T$, which represent the proportions of mosquitoes crossing the infection, dissemination, and transmission barriers, respectively; $\eta$, a boolean indicator ($\eta \in \{0,1\}$) governing the transition from state $D$ to either $T_{\text{S}}$ or $T$; $\rho$, a boolean indicator ($\rho \in \{0,1\}$) governing the transition from state $T$ to either $P_{\text{S}}$ or $P$; $\alpha_X$ and $\beta_X$, the beta distribution parameters associated with the duration in state $X \in \{I,D,T,P\}$; and $n$, $m$, $u$, and $v$, the maximum durations of stay in $I$, $D$, $T$, and $P$, respectively.
}
\label{eda1}
\end{figure*}

To study the potential cessation or intermittence of mosquito transmission capacity over time, we extended a previously developed intra-vector viral dynamics (IVD) framework \citep{loisel_intra-vector_2025} by introducing additional transmission states. The resulting model is a discrete-time stochastic mechanistic compartmental model with a quarter-day time step, directly reflecting the processes observed in vector competence experiments (Fig ~\ref{eda1}). As in the previous framework, the model includes one compartment for each major IVD stage. Stage $E$ represents exposed mosquitoes when the virus is in the digestive tract following a blood meal. Stage $I$ corresponds to infected mosquitoes when the virus is present in the midgut. Stage $D$ denotes disseminated mosquitoes when the virus has spread to the circulatory system. Stage $T$ refers to transmitter or infectious mosquitoes, marked by the presence of the virus in their saliva. To allow for dynamic patterns of transmission capacity, we further extended this transmission phase by introducing additional compartments describing permanent transmission ($T_S$), temporary interruption of transmission ($P_1$ to $P_v$), or permanent interruption of transmission ($P_S$). All processes are modelled stochastically using successive binomial draws.

\medskip
\noindent\textbf{Transmission dynamics: stable transmission, intermittence, and cessation}

To investigate alternative hypotheses regarding the temporal stability of transmission, the structure of the transmission stage was modified. Once mosquitoes reach the final sub-compartment of the disseminated stage ($D$), they enter the transmission stage. At this point, two alternative trajectories are possible, governed by parameter $\eta$. A proportion $1 - \eta$ enters compartment $T_S$, representing mosquitoes that remain in the transmitter state for the remainder of their lifespan. The remaining proportion $\eta$, enters a structured transmission stage subdivided into daily sub-compartments $T_1$,...,$T_u$. Mosquitoes in these compartments are actively transmitting but may subsequently exit this state. Upon reaching the final $T$ sub-compartment, two outcomes are possible, governed by parameter $\rho$: a proportion $1 - \rho$ transitions to $P_S$, representing permanent cessation of transmission, and a proportion $\rho$ transitions to stage P, subdivided into sub-compartments ($P_1$,...,$P_v$) representing a temporary non-transmitting phase. Mosquitoes completing stage $P$ re-enter stage $T$, generating intermittent transmission dynamics. Different transmission regimens were defined by fixing parameters $\eta$ and $\rho$ to specific values.
We compared three alternative transmission assumptions:
\begin{enumerate}
\item 	Permanent transmission model, in which mosquitoes remain transmitters for life once the transmission stage is reached $(\eta = 0, \rho = 0)$ : $EIDT$ model;
\item 	Permanent interruption of transmission, allowing permanent cessation of transmission after entry into the transmitting stage $(\eta = 1, \rho = 0)$ : $ EIDTP$ model;
\item 	Temporary interruption of transmission model, allowing reversible transitions between transmitting and non-transmitting states $(\eta = 1, \rho = 1)$ : $EIDTPT$ model.
\end{enumerate}

\medskip
\noindent\textbf{Barrier progression and stage duration dynamics}

Progression through the infection stages is governed by three parameters, $\gamma_I$, $\gamma_D$, and $\gamma_T$, representing the proportions of mosquitoes in which the infection, dissemination, and transmission barriers are overcome. Mosquitoes in which progression stops are retained in $Is$ and $Ds$ (‘s’ for stop), denoting mosquitoes in which the infection barrier but not the dissemination barrier, and the dissemination barrier but not the transmission barrier, had been crossed, respectively.

To model the distribution of residence times within stages, compartments $I$, $D$, $T$, and $P$ are subdivided into sequential daily sub-compartments. The maximum duration in stage $I$ is set equal to the experimental duration. The maximum durations in stages $D$, $T$, and $P$ are set to the experimental duration minus one, two, and three time steps, respectively, reflecting the earliest possible biological entry into these stages. When a mosquito progresses to a new stage ($I$, $D$, $T$, or $P$), it is randomly allocated among the corresponding sub-compartments using a multinomial distribution. The allocation probabilities define the residence time distribution in that stage and are derived from a beta distribution (parameters $\alpha$ and $\beta$), defined on $[0,1]$ and discretized to match the maximum stage duration. Sojourn-time probabilities are obtained from the discretized cumulative distribution functions, ensuring flexible and biologically realistic stage duration patterns.

\medskip
\noindent\textbf{Modelling the observation process}

Because mosquitoes are destructively sampled at each DPE during vector competence experiments and due to their mortality during experiments, the number of individuals processed varies across time. To reproduce this experimental design, we incorporated an observation process in which, at each DPE, a fixed number of mosquitoes ($N_{tot}$), matching the experimental sample size, is drawn from the simulated population using multinomial sampling proportional to the numbers of individuals in compartments $E$, $I$, $D$, $T$, and $P$. Sampled mosquitoes are removed from the simulated population, ensuring consistency with the destructive nature of vector competence assays.
Because post-transmission states ($P$) are not directly observable experimentally, simulated mosquitoes in state P were aggregated with disseminated mosquitoes ($D$).
Each stochastic replicate therefore represents one simulated vector competence experiment, producing stage-specific mosquito counts at each DPE directly comparable to laboratory observations.

\subsection{Parameter inference}
Model parameters were estimated using an Approximate Bayesian Computation (ABC) approach with a Sequential Monte Carlo (SMC) sampler, following the framework described in \cite{loisel_intra-vector_2025}. Briefly, ABC-SMC approximates posterior distributions by iteratively sampling parameter sets (particles), simulating data under the model, and retaining those that produce simulations sufficiently close to the observed data. Tolerance thresholds are progressively reduced across generations, allowing the particle population to converge toward the posterior distribution. The procedure was implemented using the R package BRREWABC.
For each model, the barriers and distribution of duration parameters were inferred. For all models, the proportions governing barrier progression $(\gamma_I$, $\gamma_D$, $\gamma_T)$ and the parameters $(\alpha_I, \beta_I)$ and $(\alpha_D, \beta_D)$ describing residence time distributions in the infected ($I$) and disseminated ($D$) stages were inferred. For models allowing permanent or intermittent loss of transmission, we additionally inferred the parameters $(\alpha_T, \beta_T)$ and $(\alpha_P, \beta_P)$ governing the residence time distribution in the transmitter stage ($T$), and in the intermittent stage ($P$) when applicable. Residence time distributions were modelled using beta distributions parametrized by $(\alpha, \beta)$, with prior ranges defined to allow flexible distributional shapes. Uniform prior distributions were assigned to all inferred parameters (see Table~\ref{parameters_table}).
The summary statistics used for inference corresponded to the numbers of mosquitoes in each observable stage at each DPE. For each particle, we computed stage-specific distances defined as the sum of squared differences between simulated and observed mosquito counts across all DPE. Because post-transmission states are not directly observable in experimental assays, mosquitoes simulated in state $P$ were aggregated with mosquitoes in state $D$ when computing distances. A particle was accepted if its distances were below the generation-specific tolerance thresholds. Thresholds were updated iteratively based on the 90th percentiles of the accepted particles from the previous generation. Each generation retained 1000 accepted particles, and the algorithm was run for up to 30 generations or until convergence or the stopping criteria were met.

\begin{table*}[]
\centering
\caption{Estimated model parameters and their prior distributions}
\begin{tabular}{p{1.8cm} p{5.2cm} c p{5.8cm}}
\toprule
Parameter & Description & Prior distribution used & Justification \\ \midrule

$\gamma_I$ & Proportion of mosquitoes for which the infection barrier is crossed &
$\mathcal{U}(0,1)$ &
Natural bounds of a proportion \\

$\gamma_D$ & Proportion of mosquitoes for which the dissemination barrier is crossed &
$\mathcal{U}(0,1)$ &
Natural bounds of a proportion \\

$\gamma_T$ & Proportion of mosquitoes for which the transmission barrier is crossed &
$\mathcal{U}(0,1)$ &
Natural bounds of a proportion \\

$\alpha$ & Beta distribution shape parameter &
$\mathcal{U}(0.001,100)$ &
Bounds chosen to allow a wide range of Beta distribution shapes while avoiding numerical instability near 0 \\

$\beta$ & Beta distribution shape parameter &
$\mathcal{U}(1,100)$ &
Bounds chosen to allow a wide range of Beta distribution shapes while excluding values below 1 that produce singular behavior \\

\bottomrule
\end{tabular}
\label{parameters_table}
\end{table*}

\subsection{Analyses and models comparisons}

The relevance of allowing mosquitoes to exit the transmitter ($T$) state was assessed by comparing competing epidemiological models fitted using the ABC-SMC procedure. To this end, the ability of model simulations to reproduce the observed counts across health states and observation times was evaluated. Model fit was quantified by the root mean square error (RMSE) between simulated and observed values.

For each retained ABC particle, multiple stochastic trajectories were simulated to account for process variability. This produced a distribution of RMSE values across particles reflecting posterior uncertainty in model predictions.

Model comparison was based on two complementary criteria: (i) the difference in weighted mean RMSE between models, and (ii) the probability that one model produces lower RMSE than another.
These measures provide interpretable summaries of the posterior predictive distributions of model fit.

\subsubsection*{Computation of particle-level RMSE}

Let $y_{t,s}$ denote the observed number of individuals in health state $s$ at observation time $t$, and $\hat{y}^{(p,k)}_{t,s}$ the corresponding simulated value obtained from particle $p$ and stochastic replicate $k$.

For each particle $p$ and replicate $k$, the RMSE was computed across all observation times and states as

\begin{equation}
RMSE_{p,k} =
\sqrt{
\frac{1}{TS}
\sum_{t=1}^{T}
\sum_{s=1}^{S}
\left(
\hat{y}^{(p,k)}_{t,s} - y_{t,s}
\right)^2
}
\end{equation}

where $T$ is the number of observation times and $S$ the number of health states.

To integrate stochastic variability in the model dynamics, RMSE values were averaged across the $K$ (=10) simulated trajectories for each particle:

\begin{equation}
RMSE_p =
\frac{1}{K}
\sum_{k=1}^{K} RMSE_{p,k}.
\end{equation}

This yielded one RMSE value per particle.

\subsubsection*{Weighted mean RMSE}

Particles obtained from the ABC-SMC algorithm are associated with weights $w_p$ approximating the posterior distribution of model parameters. To account for this weighting, the model-level RMSE was computed as the weighted mean across particles:

\begin{equation}
\overline{RMSE}_M =
\sum_{p=1}^{P}
w_p \, RMSE_p,
\end{equation}

where $P$ denotes the number of retained particles.
This metric summarizes the model’s ability to jointly reproduce infection, dissemination, and transmission dynamics.
Lower values of $\overline{RMSE}_M$ indicate better agreement between model simulations and observations.

\subsubsection*{Difference in RMSE between models}

For two competing models $M_i$ and $M_j$, their difference in predictive performance was quantified as

\begin{equation}
\Delta_{ij} =
\overline{RMSE}_{M_i} -
\overline{RMSE}_{M_j}.
\end{equation}

A negative value of $\Delta_{ij}$ indicates that model $M_i$ provides a better predictive fit to the data than model $M_j$, whereas a positive value indicates the opposite. Values close to zero indicate similar predictive performance.

Uncertainty in $\Delta_{ij}$ was estimated using a weighted bootstrap of particles. For each bootstrap replicate, particles were resampled with replacement according to their ABC weights and the difference in RMSE was recomputed. The resulting bootstrap distribution was used to derive 95\% confidence intervals.

\subsubsection*{Probability of dominance}

To complement the comparison based on mean RMSE, we computed the probability that a particle drawn from model $M_i$ produces a lower RMSE than a particle drawn from model $M_j$:

\begin{equation}
P_{ij} =
P(RMSE_i < RMSE_j).
\end{equation}

This probability was estimated using Monte Carlo sampling. At each iteration, one particle was drawn from each model according to their posterior weights, and the RMSE values were compared. The proportion of times where $RMSE_i < RMSE_j$ provided an estimate of $P_{ij}$.

Values of $P_{ij}$ close to $0.5$ indicate similar predictive performance, whereas values approaching $1$ indicate that model $M_i$ consistently produces lower prediction errors than model $M_j$.

\subsubsection*{Influence of transmission assumptions on posterior parameter inference}

To assess how structural assumptions regarding the transmitting state influence parameter inference, we compared posterior parameter distributions across competing models. Specifically, we examined differences between models assuming permanent transmission and those allowing transmission interruption with respect to the estimated barrier-crossing parameter $\gamma_{IDT} = \prod_{X \in \{I,D,T\}} \gamma_X$ and the duration of the infected, disseminated, and transmitting states. We also studied the distribution of duration in state $P$ for models assuming temporary interruption of transmission.

\subsection{Implementation and reproducibility}

\subsubsection*{Software and programming languages used}

All analyses were conducted primarily in R, with complementary post-processing scripts written in Python. Model inference relied principally on the BRREWABC package, and data processing and visualization were performed using core packages from the tidyverse, notably dplyr and ggplot2. Parallel computations were executed in a Sun Grid Engine (SGE) environment.

\subsubsection*{Code and data availability}

The code used to perform the analyses and generate the results presented in this study is available from the project GitHub repository: [Repository name], available at: \url{https://forge.inrae.fr/dynamo/vbd/ivd-eidtp} ([accessed 27 March 2026]). Input data required to reproduce the main analyses are provided in the repository.

\newpage
\section{RESULTS}
\label{sec:res}

\subsection{Allowing interruption of transmission improves model fit in experimental conditions exhibiting declining transmitter proportions}

Across all experimental conditions, we observed a decrease of the proportion of transmitter vectors in a subset of ten experiments (1, 3, 4, 6, 7, 8, 9, 30, 34, 49). The proportion of transmitting mosquitoes declined at later time points while the proportion of disseminated mosquitoes increased. Such non-monotonic patterns suggest a possible loss of transmission capacity over time. We therefore examined whether allowing interruption of the transmitting state improved model fit in these specific dynamical contexts. Visual inspection of posterior simulated trajectories for infected ($I$), disseminated ($D$), and transmitter ($T$) states showed a better agreement between observed and predicted dynamics when interruption was included for several experimental conditions (see Fig.~\ref{visual_fit} for experimental conditions 1 and 9, and others experimental conditions in Supplementary Figs.~\ref{visual_fit_supp}). Quantitative comparison of model-level RMSE associated to each model further supported this observation (Table~\ref{rmse_table} and Fig.~\ref{rmse_distributions} in Supplementary). Among the ten experimental conditions analyzed, the lowest model-level RMSE was obtained for a model incorporating transmission interruption in seven cases (1, 3, 4, 6,7,9, 30), with clear superiority of these models for experimental conditions 1, 3, 4, 6, and 9 (probability of having a lower model-level RMSE compared to the EIDT model > 0.6; Table~\ref{rmse_table} and supplementary Table~\ref{pairwise_rmse}). In contrast, three experimental conditions (8, 34, 49) were better explained under the assumption of permanent transmission, with clear superiority of this model for experimental conditions 34 and 49 (probability of having a lower model-level RMSE compared to the EIDTP or EIDTPT models > 0.6; Table~\ref{rmse_table} and supplementary Table ~\ref{pairwise_rmse}).

Distinguishing between permanent and temporary interruption models remained difficult, as differences in model-level RMSE between these two models were generally small (Table~\ref{rmse_table}). The probability of one model having a lower RMSE than the other was consistently below 0.6, except for experimental condition 1, for which the EIDTP model showed clear superiority over the EIDTPT model (Table~\ref{pairwise_rmse}).

The present analysis focused on a subset of 10 experimental conditions exhibiting a decline in the proportion of transmitting mosquitoes over time. In the 42 remaining experimental conditions, such patterns were not observed, either because the proportion of transmitting mosquitoes remained stable or because very few individuals reached the transmitting state, making these data less informative for assessing transmission interruption. In these cases, models without transmission interruption generally provided comparable or better fits, as reflected by the probability of obtaining a lower model-level RMSE (Table~\ref{pairwise_rmse_42} in Supplementary). In such contexts, the simpler model structure may therefore be preferred.

Overall, these results indicate that incorporating transmission interruption is not always required but appears relevant in experimental settings showing patterns consistent with a decline in transmission capacity over time.

\begin{figure*}[hbt!]
\centering
\includegraphics[width=\textwidth]{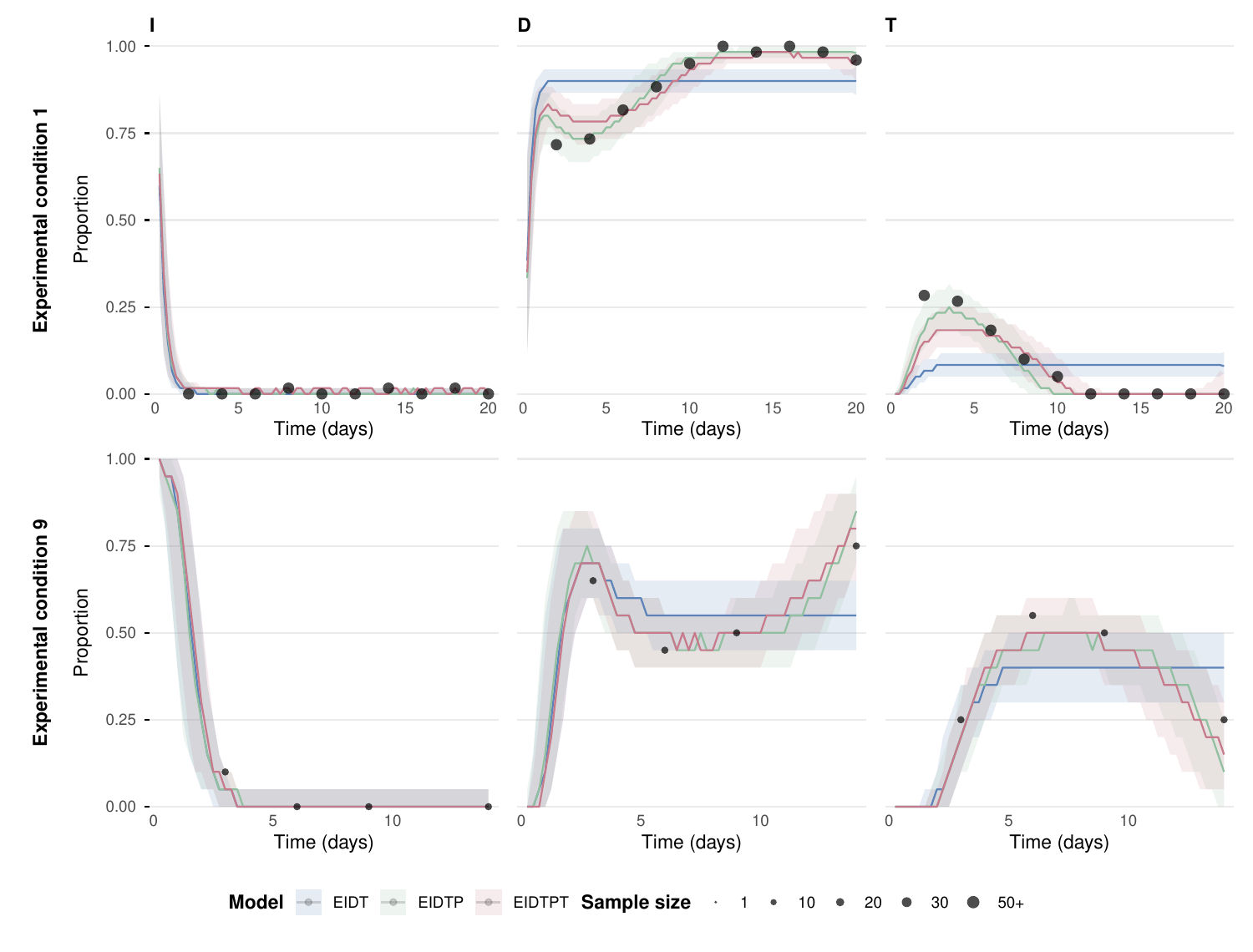}
\caption{\textbf{Comparison of visual fit between the three models (EIDT): permanent transmission, (EIDTP): permanent interruption of transmission, and (EIDTPT): temporary interruption of transmission.}
Observed and simulated dynamics of the proportions of infected ($I$), disseminated ($D$), and transmitter ($T$) mosquitoes over time for two representative experimental conditions 1 and 9 for the three models.
Points represent observed mosquitoes proportions at each sampling time, with point size proportional to the sample size. The solid line shows the median, and the shaded envelope indicates the uncertainty band spanning the 25th to the 75th percentile.}
\label{visual_fit}
\end{figure*}

\begin{table}[]
\caption{Quantitaive comparison of adequation between observed and simulated dynamics for the three models (EIDT, EIDTP, EIDTPT) .
Model-level RMSE is indicated for the three models across experimental conditions. The lowest model-level RMSE within each experimental condition is highlighted in grey.
* indicates experimental conditions where the best interruption model (EIDTP or EIDTPT) had a probability > 0.6 of yielding a lower model-level RMSE than the no-interruption model (EIDT), based on pairwise model-level RMSE comparisons.
** indicates experimental conditions where the EIDT model had a probability > 0.6 of yielding a lower model-level RMSE than the best interruption model.}
\label{rmse_table}
\begin{tabular}{p{2cm}ccc}
\toprule
Experimental condition & EIDT                          & EIDTP                         & EIDTPT                        \\ \midrule
1        & 5,079                         & \cellcolor[HTML]{C0C0C0}3,005* & 3,520                         \\
3        & 4,992                         & 3,573                         & \cellcolor[HTML]{C0C0C0}3,524* \\
4        & 2,913                         & \cellcolor[HTML]{C0C0C0}2,460* & 2,470                         \\
6        & 4,730                         & \cellcolor[HTML]{C0C0C0}3,962* & 4,050                         \\
7        & 3,522                         & 3,559                         & \cellcolor[HTML]{C0C0C0}3,520 \\
8        & \cellcolor[HTML]{C0C0C0}2,576 & 2,624                         & 2,662                         \\
9        & 2,658                         & 2,379                         & \cellcolor[HTML]{C0C0C0}2,355* \\
30       & 2,523                         & \cellcolor[HTML]{C0C0C0}2,449 & 2,506                         \\
34       & \cellcolor[HTML]{C0C0C0}2,904** & 3,351                         & 3,873                         \\
49       & \cellcolor[HTML]{C0C0C0}2,555** & 2,735                         & 2,676                         \\ \bottomrule
\end{tabular}
\end{table}

\subsection{Variation of barrier-crossing parameters under alternative assumptions on the transmitting state}

To assess how model structure influences parameter inference, we focused on experimental conditions for which allowing interruption of the transmitting state resulted in a clear improvement in model fit compared to the permanent transmission assumption (experimental conditions 1, 3, 4, 6, 9).

\begin{figure*}[hbt!]
\centering
\includegraphics[width=\textwidth]{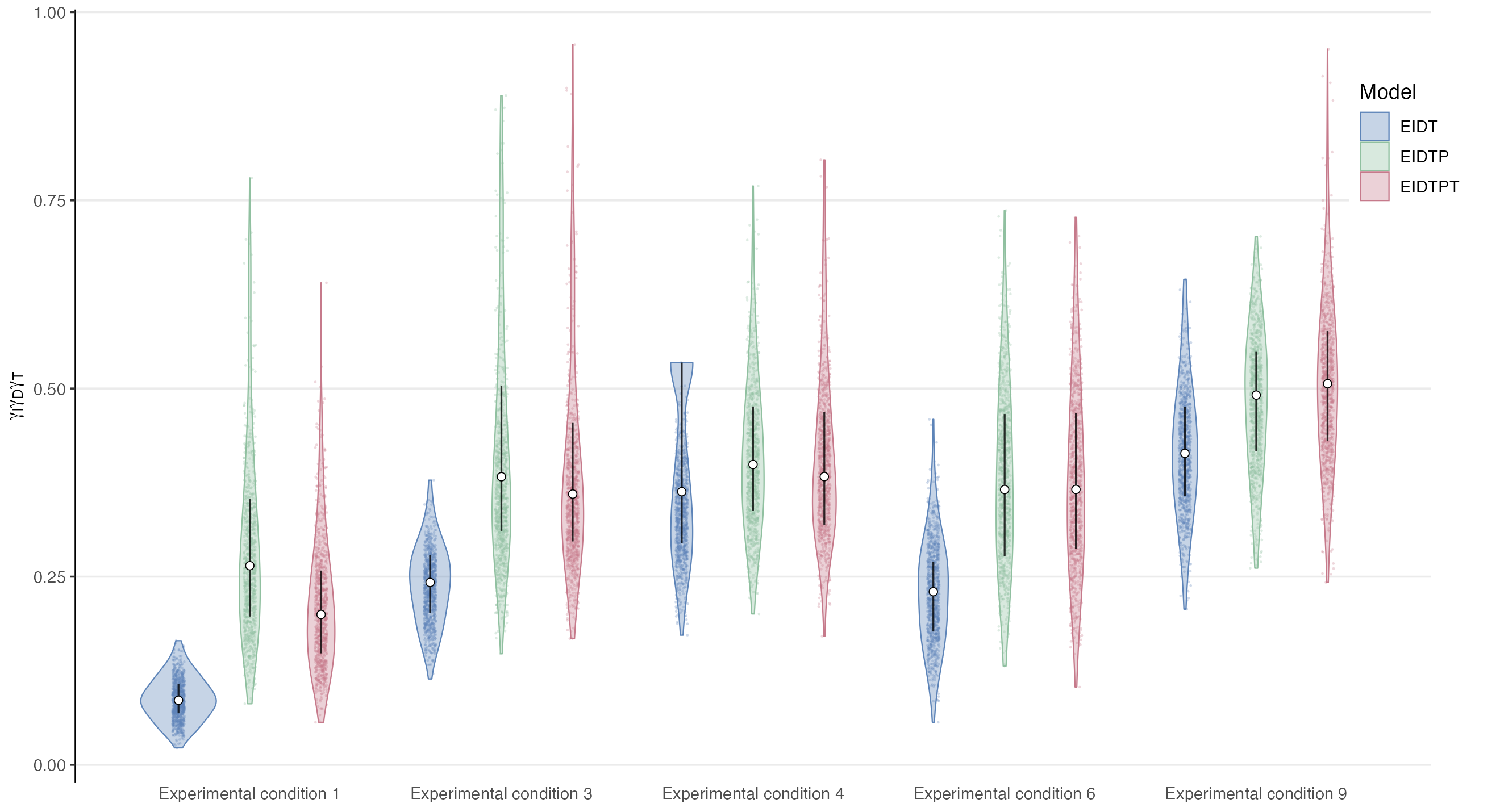}
\caption{\textbf{Posterior distributions of the barrier-crossing parameter across selected experimental conditions (scenarios) 1, 3, 4, 6 and 9 and model structures.}. Violin plots represent the posterior density of $\gamma_{IDT}$ for each experimental condition and model (EIDT: permanent transmission; EIDTP: permanent interruption of transmission; EIDTPT: temporary interruption of transmission), with white dots indicating the median and vertical bars the interquartile range (q25–q75).}
\label{violin_plot_gammaIDT}
\end{figure*}

Across this subset of experimental conditions, the posterior of $\gamma_{IDT}$ was consistently higher under models allowing interruption than under the permanent transmission assumption (delta between median of $\gamma_{IDT}$ for EIDT and EITP/EITPT models ranging from 0.02 to 0.18, see Fig~\ref{violin_plot_gammaIDT}). Because $\gamma_{IDT}$ represents the proportion of mosquitoes for which the infection, dissemination, and transmission barriers were successfully crossed, lower values correspond to stronger effective intra-mosquito barriers. The systematically lower estimates of $\gamma_{IDT}$ under the permanent transmission assumption may indicate that the inference compensates for the model’s structural restriction (i.e., the inability to leave the transmitting state) by attributing greater apparent influence to internal barriers.

In contrast, no marked differences were observed in the estimated mean duration of the infected and disseminated states between models allowing interruption and those assuming permanent transmission (see Figs.~\ref{violin_plot_betaI_mean} and ~\ref{violin_plot_betaD_mean} in Supplementary).

\subsection{Distribution of transmitting-state duration under interruption models}

We observed substantial variation in the duration of the transmitting state ($T$) across experimental conditions for which transmission interruption is relevant (experimental conditions 1, 3, 4, 6, and 9), with the mode of the mean duration ranging from approximately 5 to 10 days (Fig.~\ref{mean_duration_T} and ~\ref{duration_in_IDTP_detailed} in Supplementary). This variability reflects differences in the duration of transmission potential across experimental conditions.
In particular, experimental conditions in which the proportion of individuals capable of transmitting declines early (with a marked decrease before 10 DPE, corresponding to experimental conditions 1 and 3) also exhibit a shorter mean duration in the $T$ state, ranging from 5.5 to 6 days, compared with 9 to 11.6 days in the other experimental conditions.

\begin{figure}[hbt!]
\centering
\includegraphics[width=\textwidth]{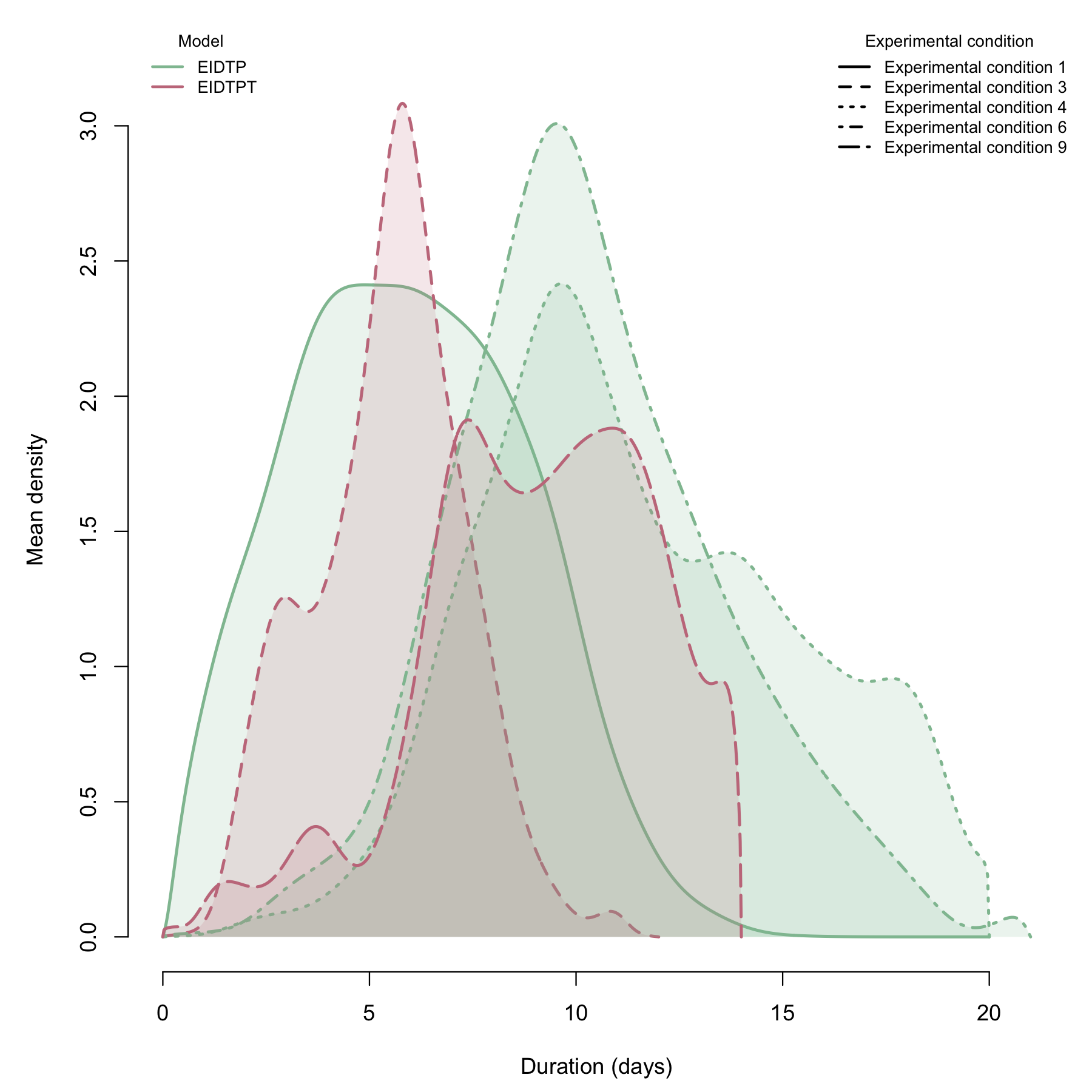}
\caption{\textbf{Distribution of transmitting-state duration across experimental conditions under models allowing transmission interruption. }
Posterior distributions of the duration spent in the transmitting state ($T$) across  experimental conditions 1, 3, 4, 6, and 9, estimated under models allowing transmission interruption (EIDTP or EIDTPT). For each experimental condition, only the best-fitting model (based on level-model RMSE; see previous section) is displayed. Densities represent the distribution of the mean duration in state $T$ (in days). Colours indicate model structures, while line types distinguish experimental conditions.}
\label{mean_duration_T}
\end{figure}

In experimental conditions where the intermittent interruption model (EIDTPT) provided the best fit (experimental conditions 3 and 9), we further examined the estimated duration of the non-transmitting state ($P$). The posterior distributions of the mean duration in state $P$ showed relatively consistent patterns across these experimental conditions, with modes around 6 days (Fig.~\ref{mean_duration_P_detailed}). This suggests that, under the intermittent interruption assumption, mosquitoes may remain in a non-transmitting state for several days before potentially regaining transmission capacity.

\section{DISCUSSION}
\label{sec:disc}

By developing and comparing intra-vector viral dynamics models that differ in whether mosquitoes can leave the transmitting state, we showed that this mechanism is likely relevant in vector competence experiments where the proportion of transmitting mosquitoes declines over time. In such experiments, accounting for this mechanism led to higher estimates of the proportion of mosquitoes that reach the transmitting state. This finding challenges the common assumption that once a mosquito reaches the transmitting state, it remains infectious for the rest of its life \citep{robison_comparison_2020}.

This work provides a potential mechanistic explanation for patterns observed in some vector competence experiments \citep{robison_comparison_2020,merwaiss_chikungunya_2021,prudhomme_native_2019,amraoui_culex_2016,seixas_potential_2018}. Although our results do not allow us to conclusively demonstrate that the observed decline in transmitting mosquitoes is caused by a true interruption of transmission capacity, they support the plausibility of this hypothesis and highlight the need for further investigation. Alternatively, the apparent loss of transmission capacity could reflect fluctuations of viral load around the detection threshold of saliva detection assays. Indeed experiment on Aedes albopictus infected with CHIKV have shown a decrease in viral load in saliva over time \citep{viginier_chikungunya_2023-1}. Another explanation could be a failure to detect the virus due to the absence of saliva expectoration during sampling, rather than a true biological interruption of transmission \citep{prudhomme_native_2019}.

The possibility that mosquitoes may cease to be infectious could have important implications for arbovirus transmission dynamics. If mosquitoes can lose their transmission capacity over time, the effective duration of infectiousness may be shorter than commonly assumed in epidemiological models, which typically consider mosquitoes infectious for the remainder of their lifespan once the saliva is infected \citep{andraud_dynamic_2012,reiner_systematic_2013}. This could lead to overestimation of transmission potential. It would therefore be valuable to couple this extended IVD framework with a vector–host transmission model in order to evaluate how this phenomenon might influence epidemic dynamics in host populations.

Allowing mosquitoes to exit the transmitting state in the IVD model improved the agreement between simulated and observed data, particularly for chikungunya virus. In our previous work \citep{loisel_intra-vector_2025}, we showed that intra-vector viral dynamics are generally much faster for CHIKV than for DENV and ZIKV. Across experimental conditions, the timing of first observed transmission suggests that mosquitoes often reach the transmitting state within the first few days post-exposure for CHIKV (often 3 days), whereas arrival in the transmitting state tends to occur later for DENV and ZIKV, most often around 7--14 days for DENV and 9--21 days for ZIKV, depending on the experimental conditions. Thus, within a standard 21-day vector competence experiment \citep{souza-neto_aedes_2019, diallo_dengue_2022,kain_not_2022}, CHIKV-infected mosquitoes may spend a substantial fraction of the observation period in the transmitting state, potentially leaving enough time for a subsequent decline in transmission to become visible. By contrast, for DENV and ZIKV, mosquitoes often reach the transmitting state much later, so that any comparable decline may remain difficult to detect within the usual experimental time frame. Extending the duration of vector competence experiments would therefore help determine whether similar post-transmission declines also occur for other arboviruses.

A second important finding of this study is the potential overestimation of intra-mosquito barriers when transmission interruption is not incorporated into the model structure. This likely reflects a structural constraint of the classical model, in which mosquitoes cannot exit the transmitting state. When a decline in the proportion of transmitting mosquitoes is observed in the data, the inference compensates for this pattern by reducing the estimated proportion of mosquitoes that cross the transmission barrier. More broadly, this issue may also affect the interpretation of vector competence experiments. Transmission rates are often estimated at a single terminal time point as the proportion of mosquitoes with virus detected in saliva among those tested, or among those with disseminated infection \citep{wu_minimum_2022,souza-neto_aedes_2019}. If mosquitoes can temporarily or permanently lose detectable virus in their saliva after reaching the transmitting state, such estimates may underestimate the true proportion of mosquitoes that become capable of transmission. This could bias estimates of vector competence and affect our understanding of arbovirus transmission dynamics.

Our study also has several limitations that should be acknowledged. First, although our model considers two alternative forms of transmission interruption, temporary and permanent, the available data did not allow us to clearly distinguish between these two assumptions. Posterior dynamics trajectories produced by the two models were visually very similar, and the model-level RMSE obtained under both assumptions were often very close without clear superiority of one model comparing the other (Table.~\ref{rmse_table} and Table~\ref{pairwise_rmse} in Supplementary). In the experimental data analyzed here, experimental conditions of interest generally show a decline in the proportion of transmitting mosquitoes over time without a subsequent increase. This pattern may appear more consistent with a permanent loss of transmission capacity, which could suggest that a model assuming permanent interruption of transmission might provide a better explanation of the observed dynamics than the model with temporary interruption of transmission. Therefore, the intermittent transmission model (EIDTPT) can become nearly equivalent to the permanent interruption model (EIDTP) when the estimated duration of the non-transmitting stage is long. This seems to be the case for the two experimental conditions in which the intermittent model was selected (experimental conditions 3 and 9), where the mode of the mean duration in the non-transmitting state ($P$) was around 6 days (Fig.~\ref{mean_duration_P_detailed}). Overall, these results suggest that permanent interruption of transmission may be the most parsimonious explanation for the observed dynamics. However, it remains possible that transmission could resume after a longer delay than the duration of the experiments analyzed here. Longer-term longitudinal experiments would therefore be valuable for evaluating this assumption.

Although models allowing transmission interruption clearly perform better under some experimental conditions, both visually and in terms of model-level RMSE (experimental conditions 1, 3, 4, 6, and 9), this pattern is not observed consistently across all conditions. Indeed, in several experimental conditions (7, 8, 30), model-level RMSE and simulated dynamics were very similar across models and for experimental conditions 34 and 49, the model without transmission interruption appears to perform better. In these 5 cases, the experimental conditions often correspond to situations where the decline in the proportion of transmitting mosquitoes was either of small amplitude (e.g., experimental condition 49) or observed at only a single time point (experimental conditions 8, 30, and 34), in contrast to experimental conditions where this decrease was more pronounced. Consequently, the data for these experimental conditions appear to be less informative with respect to the underlying transmission dynamics, which may explain the difficulty in clearly selecting one model structure over another.

Among the 52 experimental conditions included in the dataset, only 10 were analyzed in detail in this study because they exhibited a decrease in the proportion of mosquitoes in the transmitting state ($T$) while the proportion in the disseminated state ($D$) increased. In the remaining experimental conditions, where such patterns were not observed, the data do not provide evidence supporting the presence of transmission interruption, making the inclusion of this mechanism less relevant and suggesting that a simpler model structure may be preferred. However, it is important to note that in a substantial number of experimental conditions (22 out of 52), very few or no mosquitoes reached the transmitting state, making it impossible to observe such patterns. For these experimental conditions, the data suggest that intra-mosquito barriers may be particularly strong. Increasing the initial number of mosquitoes in vector competence experiments could therefore be useful in order to obtain, by the end of the experiment, a sufficient number of mosquitoes reaching the transmitting state. This would allow a more reliable investigation of transmission dynamics in this final stage.

Increasing the number of mosquitoes reaching the transmitting state would provide more informative data to better characterize not only the temporal dynamics of transmission, but also the heterogeneity in the transmission trajectories followed by individual mosquitoes. In our study, the values of the parameters $\rho$ and $\eta$, which control the proportions of mosquitoes entering the different possible transmission states, were fixed in order to facilitate comparison between alternative assumptions and to avoid potential estimation issues arising from the limited number of mosquitoes reaching the transmitting state. Allowing these parameters to vary could have captured inter-individual variability in antiviral responses among mosquitoes, which has previously been documented \citep{raquin_individual_2017}.

Overall, our results suggest that the classical assumption of lifelong infectiousness in mosquitoes may not always hold, and that incorporating dynamic transmission states into intra-vector models could improve the interpretation of vector competence experiments. Further experimental studies designed to monitor transmission dynamics over time will be essential to determine the biological mechanisms underlying these patterns.

Notably, presence of infectious virus and viral load could be monitored daily in various compartments of individual mosquitoes across their whole lifetime, although this relies on the design of non-lethal sampling techniques. Recent tools such as vectorchip \citep{kumar_microfluidic_2021}, a microfluidic platform for molecular interrogation of mosquito bites, or virus detection from mosquito excreta \citep{fontaine_excretion_2016} opens new avenues for non-invasive individual monitoring of mosquito infection that will likely improve the modeling of arbovirus transmission potential by mosquitoes.

\begin{acknowledgement}
We are grateful to the INRAE MIGALE bioinformatics facility (MIGALE, INRAE, 2020. Migale bioinformatics Facility, doi: 10.15454/1.5572390655343293E12) for providing help and/or computing and/or storage resources.
\end{acknowledgement}

\begin{authorcontributions}
GB and LL conceived and designed the study. Data curation was performed by LL and TM. The methodology was developed by GB and LL. Software implementation and the development of supporting algorithms were carried out by GB, LL, and TM. GB, LL, and TM analyzed the results. Visualizations were prepared by GB and LL. The original draft of the manuscript was written by LL and GB. Review and editing were performed by LL, GB, PE, VR and TM.
\end{authorcontributions}


\bibliography{biblio}

\appendix
\clearpage
\onecolumn

\renewcommand{\thefigure}{S\arabic{figure}}
\setcounter{figure}{0}

\renewcommand{\thetable}{S\arabic{table}}
\setcounter{table}{0}

\begin{sidewaystable}[hbt!]
\centering
\caption{Summary of experimental data used to infer the three IVD model for chikungunya virus (CHIKV). ID:Infectious Dose (Log10 FFU/mL), FFA : Fluorescent Focus Assay, RT\-PCR:reverse transcription polymerase chain reaction Dpe: Day post exposition, ND:not defined }
\label{chikv_table}
\resizebox{\textwidth}{!}{%

\begin{tabular}{@{}lllllllllll@{}}
\toprule
\begin{tabular}[c]{@{}l@{}} Experimental conditions \\ \end{tabular} & Virus species & Experimental condition name  & \begin{tabular}[c]{@{}l@{}}Virus strain \\ (origin)\end{tabular} & \begin{tabular}[c]{@{}l@{}}Mosquito genus\\ and species \\ (origin)\end{tabular} & ID  & \begin{tabular}[c]{@{}l@{}}Ambiance conditions: \\ Temperature (°c) \\ Humidity (\%)  \\ Light-dark cycle (hours)\end{tabular} & Dpe & \begin{tabular}[c]{@{}l@{}}Mosquito number \\ (mean by Dpe)\end{tabular} & \begin{tabular}[c]{@{}l@{}}Mosquitoes parts \\ analyzed for:\\ Infection (I) \\ Dissemination(D) \\ Transmission (T)\\ (method)\end{tabular} & Ref  \\ \midrule
1 & CHIKV & BriVirIsl\_aeg\_PozRic\_6.9 & 99659  & Ae.aegypti & 6.9 & 28°c & 2,4,6,8,10,12,14,16,18,20 & 60 & I:midgut(RT-PCR) & \citep{robison_comparison_2020} \\
  & & & (British Virgin Islands)  & \multicolumn{2}{l}{(Mexico, Poza Rica)} & 70-80 &  &  & \multicolumn{2}{l}{D:legs/wings(RT-PCR)} \\
  &  & & & & & 12h:12h &  &  & \multicolumn{2}{l}{T:saliva (PFA)}  \\
2 & CHIKV  & CarIls\_2013\_aeg\_Guy\_6 &Carribean strain-2013  & Ae.aegypti & 6 & 28°c & 3,5,7,10,14  & 41 & I:body(PFA) & \citep{wang_current_2022}  \\
  & & & (Carribean Island)  & \multicolumn{2}{l}{(French Guiana)} & 80 &  &  & \multicolumn{2}{l}{D:head(PFA)} \\
  &  & & & & & 12h:12h &  &  & saliva(PFA) &  \\
3 & CHIKV  & FrCarIs\_aeg\_Tha\_6 & Carribean strain & Ae.aegypti & 6 & 28°c & 3,6,9,12 & 48 & I:body(PFA) & \citep{merwaiss_chikungunya_2021} \\
  & & & (French carribean Island) & \multicolumn{2}{l}{(Thailand, KPP)} & 70 &  &  & \multicolumn{2}{l}{D:head(PFA)} \\
  &  & & & & & 12h:12h &  &  & saliva(PFA) &  \\
4 & CHIKV  & Ind\_albo\_Tir\_8  &  06.21  & Ae.albopictus & 8 & 28 +/-1°c  & 3,5,7,10,12,14,20  & 18 & I:body(FFA) & \citep{prudhomme_native_2019} \\
  & & & (India)  & \multicolumn{2}{l}{(Albania, Tirana)} & 80 &  &  & \multicolumn{2}{l}{D:head(FFA)} \\
  &  & & & & & 16h:8h &  &  & \multicolumn{2}{l}{T:saliva(FFA)}  \\
5 & CHIKV  & Ind\_gen\_Tir\_8 &  06.21  & Ae.geniculatus & 8 & 28 +/-1°c  & 3,5,7,10,12,14,20  & 19 & I:body(FFA) & \citep{prudhomme_native_2019} \\
  &  & & (India)  & \multicolumn{2}{l}{(Albania, Tirana)} & 80 &  &  & \multicolumn{2}{l}{D:head(FFA)} \\
  &  & & & & & 16h:8h &  &  & \multicolumn{2}{l}{T:saliva(FFA)}  \\
6 & CHIKV  & LaR\_albo\_Rab\_7 & 06.21  & Ae.albopictus & 7 & 28 +/- 1°c  & 3,7,14,21 & 30 & I:abdomen/thorax(FFA)  & \citep{amraoui_potential_2019} \\
  & & & (Reunion Island) & \multicolumn{2}{l}{(Morocco, Rabat)} & 80 &  &  & \multicolumn{2}{l}{D:head (FFA)}  \\
  &  & & & & & 16h:8h &  &  & \multicolumn{2}{l}{T:saliva (FFA)}  \\
7 & CHIKV & LaR\_albo\_Tun\_7 & 06.21  & Ae.albopictus & 7 & 28 +/- 1°c  & 3,7,10,14,21  & 27 & I:abdomen (FFA) & \citep{bohers_recently_2020} \\
  & & & (Reunion Island) & \multicolumn{2}{l}{(Tunisia,Car, Ami, La Mar)}  & 80 &  &  & \multicolumn{2}{l}{D:thorax/head (FFA)}  \\
  &  & & & & & 16h:8h &  &  & \multicolumn{2}{l}{T:saliva (FFA)}  \\
8 & CHIKV & Ncal\_aeg\_Mad\_Fun\_7.3  & NC/2011-568 & Ae.aegypti & 7.3 & 28 +/- 1°c  & 3,6,9,14 & 20 & I:thorax/abdomen (ND)  & \citep{seixas_potential_2018} \\
  & & & (New Caledonia) & \multicolumn{2}{l}{(Mad Island, Funchal)}  & 80 &  &  & I:head(FFA) &  \\
  &  & & & & & ND &  &  & \multicolumn{2}{l}{T:saliva(FFA)}  \\
9 & CHIKV & Ncal\_aeg\_Mad\_Pa\_do\_Ma\_7.3  & NC/2011-568 & Ae.aegypti & 7.3 & 28 +/- 1°c  & 3,6,9,14 & 20 & I:thorax/abdomen (ND)  & \citep{seixas_potential_2018} \\
  & & & (New Caledonia) & \multicolumn{2}{l}{(Mad Island, P. do Mar)}  & 80 &  &  & I:head(FFA) &  \\
  &  & & & & & ND &  &  & \multicolumn{2}{l}{T:saliva(FFA)}  \\
10 & CHIKV  & Tah\_aeg\_Tah\_7 & PF14/300914-109 & Ae.aegypti & 7 & 27°c & 6,9,14,21 & 39 & I:thorax/abdomen (RT-PCR)  & \citep{richard_vector_2016} \\
  &  & & (Tahiti)  & \multicolumn{2}{l}{(Tahiti island, Toahotu)}  & 80 &  &  & \multicolumn{2}{l}{D:legs (RT-PCR)}  \\
  &  & & & & & 12h:12h &  &  & \multicolumn{2}{l}{T:saliva (FFA)}  \\ \bottomrule
\end{tabular}
}

\end{sidewaystable}

\begin{sidewaystable}[]
\centering
\caption{Summary of experimental data used to infer the three IVD model for dengue virus (DENV). ID:Infectious Dose (Log10 FFU/mL), FFA : Fluorescent Focus Assay, RT\-PCR:reverse transcription polymerase chain reaction Dpe: Day post exposition, ND:not defined }
\resizebox{\textwidth}{!}{%
\begin{tabular}{@{}lllllllllll@{}}
\toprule
\begin{tabular}[c]{@{}l@{}} Experimental conditions \\  \end{tabular} & Virus species & Experimental condition name  & \begin{tabular}[c]{@{}l@{}}Virus strain \\ (origin)\end{tabular} & \begin{tabular}[c]{@{}l@{}}Mosquito genus \\ and species\\ (origin)\end{tabular} & ID & \begin{tabular}[c]{@{}l@{}}Ambiance conditions: \\ Temperature (°c)\\ Humidity (\%)  \\ Light-dark cycle (hours)\end{tabular} & Dpe  & \begin{tabular}[c]{@{}l@{}}Mosquito number \\ (mean by Dpe)\end{tabular} & \begin{tabular}[c]{@{}l@{}}Mosquitoes parts analyzed for \\ Infection (I) \\ Dissemination(D) \\ Transmission (T)  \\ (method)\end{tabular} & Ref \\ \midrule
11 & DENV-1 & Sing\_albo\_Mon\_5  & SG (EHI)D1/30889Y14  & Ae.albopictus & 5 & ND  & 7,14,21,28 & 18 & I:body(PFA) & \citep{fortuna_assessing_2024} \\
  &  &  & (Singapore) & \multicolumn{2}{l}{(Italia, Montecchio)}  & ND  & &  & \multicolumn{2}{l}{D:legs/wings(PFA)}  \\
  &  &  & & & & ND  & &  & \multicolumn{2}{l}{T:saliva (PFA)} \\
12 & DENV-1 & Sing\_albo\_Rey\_5  & SG (EHI)D1/30889Y14  & Ae.albopictus & 5 & ND  & 7,14,21,28 & 17 & I:body(PFA) & \citep{fortuna_assessing_2024} \\
  &  &  & (Singapore) & \multicolumn{2}{l}{(Italia, Reynosa)} & ND  & &  & \multicolumn{2}{l}{D:legs/wings(PFA)}  \\
  &  &  & & & & ND  & &  & \multicolumn{2}{l}{T:saliva (PFA)} \\
13 & DENV-1 & Sing\_albo\_Rom\_5  & SG (EHI)D1/30889Y14  & Ae.albopictus & 5 & ND  & 7,14,21,28 & 18 & I:body(PFA) & \citep{fortuna_assessing_2024} \\
  &  &  & (Singapore) & \multicolumn{2}{l}{(Italia, Rome)} & ND  & &  & \multicolumn{2}{l}{D:legs/wings(PFA)}  \\
  &  &  & & & & ND  & &  & \multicolumn{2}{l}{T:saliva (PFA)} \\
14 & DENV-2 & Bang\_albo\_Rab\_7  & Prof Leon Rosen & Ae.albopictus & 7 & 28 +/- 1°c & 3,7,14,21  & 28 & I:abdomen/thorax (FFA) & \citep{amraoui_potential_2019} \\
  &  &  & (Thailand, Bangkok)  & \multicolumn{2}{l}{(Morocco, Rabat)} & 80  & &  & \multicolumn{2}{l}{D:head (FFA)} \\
  &  &  & & & & 16h:8h & &  & \multicolumn{2}{l}{T:saliva (FFA)} \\
15 & DENV-2 & Bang\_albo\_Tun\_7  & (Thailand, Bangkok)  & Ae.albopictus & 7 & 28 +/- 1°c & 3,7,10,14,21 & 28 & I:abdomen (FFA)  & \citep{bohers_recently_2020} \\
  &  &  & & \multicolumn{2}{l}{(Tunisia,Car, Ami, La Mar)}  & 80  & &  & \multicolumn{2}{l}{D:thorax/head (FFA)}  \\
  &  &  & & & & 16h:8h & &  & \multicolumn{2}{l}{T:saliva (FFA)} \\
16 & DENV-3 & Cair\_1998\_aeg\_Cair\_4.7 & 1998- GenBank JN406514  & Ae aegypti & 4.7 & 28°c & 2,3,4,5,6,7,10,14 & 21 & I:body(ELISA)  & \citep{ritchie_explosive_2013}\\
  &  &  & & \multicolumn{2}{l}{(Australia,Cairns)} & 75  & &  & \multicolumn{2}{l}{D:legs/wings (ELISA)}  \\
  &  &  & & & & 12h:12h & &  & \multicolumn{2}{l}{T:saliva(ELISA)} \\
17 & DENV-3 & Cair\_2008\_aeg\_Cair\_4.9 & 2008a - GenBank JN406515  & Ae aegypti & 4.9 & 28°c & 2,3,4,5,6,7,10,14 & 21 & I:body(ELISA)  & \citep{ritchie_explosive_2013} \\
  &  &  & (Cairns)  & \multicolumn{2}{l}{(Australia,Cairns)} & 75  & &  & \multicolumn{2}{l}{D:legs/wings (ELISA)}  \\
  &  &  & & & & 12h:12h & &  & \multicolumn{2}{l}{T:saliva(ELISA)} \\ \bottomrule
\end{tabular}
}
\label{denv_table}
\end{sidewaystable}

\begin{sidewaystable}[hbt!]
\centering
\caption{Summary of experimental data used to infer the three IVD model for Rift valley fever virus (RVFV). ID:Infectious Dose (Log10 FFU/mL), FFA : Fluorescent Focus Assay, RT\-PCR:reverse transcription polymerase chain reaction Dpe: Day post exposition, ND:not defined }
\resizebox{\textwidth}{!}{%
\begin{tabular}{@{}lllllllllll@{}}
\toprule
\begin{tabular}[c]{@{}l@{}} Experimental conditions \\  \end{tabular} & Virus species & Experimental condition name  & \begin{tabular}[c]{@{}l@{}}Virus strain \\ (origin)\end{tabular} & \begin{tabular}[c]{@{}l@{}}Mosquito genus \\ and species\\ (origin)\end{tabular} & ID  & \begin{tabular}[c]{@{}l@{}}Ambiance conditions: \\ Temperature (°c)\\ Humidity (\%)  \\ Light-dark cycle (hours)\end{tabular} & Dpe & \begin{tabular}[c]{@{}l@{}}Mosquito number \\ (mean by Dpe)\end{tabular} & \begin{tabular}[c]{@{}l@{}}Mosquitoes parts analyzed for \\ Infection (I) \\ Dissemination(D) \\ Transmission (T)  \\ (method)\end{tabular} & Ref \\ \midrule
18 & RVFV  & PAEA\_ZH548\_RGC\_ab  & ZH548  & Aedes aegypti & 5,10\textasciicircum{}6 FFU/mL & 26°C & \multicolumn{2}{l}{3,7,10,14,21} & I:abdomen/thorax(PFA) & \citep{baudon_influence_2026} \\
  &  &  & (Egypt)  & (France)  &  & 80\% &  &  & \multicolumn{2}{l}{D:head PFA)} \\
  &  &  & & &  & 12h:12h &  &  & \multicolumn{2}{l}{T:saliva (PFA)} \\
19 & RVFV  & RVF654\_CulAnt\_MorAnk\_8,35 & 654608  & Culex antennatus & 8,35  & 25  & 5,5,8,14 & 16 & I:midgut(qRT-PCR)  & \citep{nepomichene_vector_2018} \\
  &  &  & (Fianarantsoa) & \multicolumn{2}{l}{(Madagascar)} &  &  &  & \multicolumn{2}{l}{D:head/legs/wings (qRT-PCR)} \\
  &  &  & & &  &  &  &  & \multicolumn{2}{l}{T:saliva (qRT-PCR)}  \\
20 & RVFV  & RVFAnD\_AedVex\_BarSen\_6-8 & AnD 133719 & Aedes vexans & 6-8  & 27  & 5,10,15,20 & 37 & I:abdomen(RT-PCR)  & \citep{ndiaye_vector_2016} \\
  &  &  & (Mauritania) & (Senegal)  &  &  &  &  & \multicolumn{2}{l}{D:legs/wings (RT-PCR)}  \\
  &  &  & & &  &  &  &  & \multicolumn{2}{l}{T:saliva (RT-PCR)}  \\
21 & RVFV  & RVFAnD\_CulPip\_BarSen\_6-8 & AnD 133719 & Culex pipiens quinquefasciatus & 6-8  & 27  & 5,10,15,20 & 30 & I:abdomen(RT-PCR)  & \citep{ndiaye_vector_2016} \\
  &  &  & (Mauritania) & (Senegal)  &  &  &  &  & \multicolumn{2}{l}{D:legs/wings (RT-PCR)}  \\
  &  &  & & &  &  &  &  & \multicolumn{2}{l}{T:saliva (RT-PCR)}  \\
22 & RVFV  & RVFAnD\_CulPoi\_BarSen\_6-8 & AnD 133719 & Culex poicilipes & 6-8  & 27  & 5,10,15,20 & 23 & I:abdomen(RT-PCR)  & \citep{ndiaye_vector_2016} \\
  &  &  & (Mauritania) & (Senegal)  &  &  &  &  & \multicolumn{2}{l}{D:legs/wings (RT-PCR)}  \\
  &  &  & & &  &  &  &  & \multicolumn{2}{l}{T:saliva (RT-PCR)}  \\
23 & RVFV  & RVFArD\_AedVex\_BarSen\_6-8 & ArD 141967 & Aedes vexans & 6-8  & 27  & 5,10,15,20 & 28 & I:abdomen(RT-PCR)  & \citep{ndiaye_vector_2016} \\
  &  &  & (Mauritania) & (Senegal)  &  &  &  &  & \multicolumn{2}{l}{D:legs/wings (RT-PCR)}  \\
  &  &  & & &  &  &  &  & \multicolumn{2}{l}{T:saliva (RT-PCR)}  \\
24 & RVFV  & RVFArD\_CulPip\_BarSen\_6-8 & ArD 141967 & Culex pipiens quinquefasciatus & 6-8  & 27  & 5,10,15,20 & 37 & I:abdomen(RT-PCR)  & \citep{ndiaye_vector_2016} \\
  &  &  & (Mauritania) & (Senegal)  &  &  &  &  & \multicolumn{2}{l}{D:legs/wings (RT-PCR)}  \\
  &  &  & & &  &  &  &  &  & \\
25 & RVFV  & RVFArD\_CulPoi\_BarSen\_6-8 & ArD 141967 & Culex poicilipes & 6-8  & 27  & 5,10,15,20 & 42 & I:abdomen(RT-PCR)  & \citep{ndiaye_vector_2016} \\
  &  &  & (Mauritania) & (Senegal)  &  &  &  &  & \multicolumn{2}{l}{D:legs/wings (RT-PCR)}  \\
  &  &  & & &  &  &  &  & \multicolumn{2}{l}{T:saliva (RT-PCR)}  \\
26 & RVFV  & RVFSH\_AedVex\_BarSen\_6-8 & SH 172805 & Aedes vexans & 6-8  & 27  & 5,10,15,20 & 41 & I:abdomen(RT-PCR)  & \citep{ndiaye_vector_2016} \\
  &  &  & (Mauritania) & (Senegal)  &  &  &  &  & \multicolumn{2}{l}{D:legs/wings (RT-PCR)}  \\
  &  &  & & &  &  &  &  & \multicolumn{2}{l}{T:saliva (RT-PCR)}  \\
27 & RVFV  & RVFSH\_CulPip\_BarSen\_6-8 & SH 172805 & Culex pipiens quinquefasciatus & 6-8  & 27  & 5,10,15,20 & 64 & I:abdomen(RT-PCR)  & \citep{ndiaye_vector_2016} \\
  &  &  & (Mauritania) & (Senegal)  &  &  &  &  & \multicolumn{2}{l}{D:legs/wings (RT-PCR)}  \\
  &  &  & & &  &  &  &  &  & \\
28 & RVFV  & RVFSH\_CulPoi\_BarSen\_6-8 & SH 172805 & Culex poicilipes & 6-8  & 27  & 5,10,15,20 & 30 & I:abdomen(RT-PCR)  & \citep{ndiaye_vector_2016} \\
  &  &  & (Mauritania) & (Senegal)  &  &  &  &  & \multicolumn{2}{l}{D:legs/wings (RT-PCR)}  \\
  &  &  & & &  &  &  &  & \multicolumn{2}{l}{T:saliva (RT-PCR)}  \\
29 & RVFV  & SLAB\_ZH548\_RGC\_ab  & ZH548  & Culex quinquefasciatus  & 5,10\textasciicircum{}6 FFU/mL & 26°C & \multicolumn{2}{l}{3,7,10,14,21} & I:abdomen/thorax(PFA) & \citep{baudon_influence_2026} \\
  &  &  & (Egypt)  & (France)  &  & 80\% &  &  & \multicolumn{2}{l}{D:head PFA)} \\
  &  &  & & &  & 12h:12h &  &  & \multicolumn{2}{l}{T:saliva (PFA)} \\
30 & RVFV  & Vexans\_ZH548\_RGC\_ab & ZH548  & Aedes vexans & 5,10\textasciicircum{}6 FFU/mL & 26°C & \multicolumn{2}{l}{3,7,10,14,21} & I:abdomen/thorax(PFA) & \citep{baudon_influence_2026} \\
  &  &  & (Egypt)  & (France)  &  & 80\% &  &  & \multicolumn{2}{l}{D:head PFA)} \\
  &  &  & & &  & 12h:12h &  &  & \multicolumn{2}{l}{T:saliva (PFA)} \\ \bottomrule
\end{tabular}
}
\label{rvfv_table}
\end{sidewaystable}

\begin{sidewaystable}[hbt!]
\centering
\caption{Summary of experimental data used to infer the three IVD model for West-Nile virus (WNV). ID:Infectious Dose (Log10 FFU/mL), FFA : Fluorescent Focus Assay, RT\-PCR:reverse transcription polymerase chain reaction Dpe: Day post exposition, ND:not defined }
\resizebox{\textwidth}{!}{%
\begin{tabular}{@{}lllllllllll@{}}
\toprule
\begin{tabular}[c]{@{}l@{}} Experimental conditions \\ \end{tabular} & Virus species & Experimental condition name & \begin{tabular}[c]{@{}l@{}}Virus strain \\ (origin)\end{tabular} & \begin{tabular}[c]{@{}l@{}}Mosquito genus \\ and species\\ (origin)\end{tabular} & ID & \begin{tabular}[c]{@{}l@{}}Ambiance conditions: \\ Temperature (°c)\\ Humidity (\%)  \\ Light-dark cycle (hours)\end{tabular} & Dpe  & \begin{tabular}[c]{@{}l@{}}Mosquito number \\ (mean by Dpe)\end{tabular} & \begin{tabular}[c]{@{}l@{}}Mosquitoes parts analyzed for \\ Infection (I) \\ Dissemination(D) \\ Transmission (T)  \\ (method)\end{tabular} & Ref \\ \midrule
31 & WNV  & WN0195\_CulPip\_SanCru\_7.65\_27\_A & 02-1956  & Culex pipiens quinquefasciatus & 7,65 & 27°C & 7,11,14,21,28 & 42 & I : abdomen/thorax (PFA) & \citep{eastwood_west_2011} \\
  &  & & (USA)  & \multicolumn{2}{l}{Galapagos}  &  & &  & \multicolumn{2}{l}{D: legs(PFA)} \\
  &  & & & & &  & &  & \multicolumn{2}{l}{T : saliva(PFA)} \\
32 & WNV  & WN0195\_CulPip\_SanCru\_7.65\_30\_A & 02-1956  & Culex pipiens quinquefasciatus & 7,65 & 30°C & 7,11,14,21  & 40 & I : abdomen (PFA)  & \citep{eastwood_west_2011} \\
  &  & & (USA)  & \multicolumn{2}{l}{Galapagos}  &  & &  & \multicolumn{2}{l}{D: legs(PFA)} \\
  &  & & & & &  & &  & \multicolumn{2}{l}{T : saliva(PFA)} \\
33 & WNV  & WNIA0\_AedVex\_IowUSA\_7.1\_27\_A & IA02-crow & Aedes vexans & 7,1 & 27  & 7,10,14,18  & 40 & I : abdomen (PFA)  & \citep{tiawsirisup_vector_2008} \\
  &  & & (Iowa)  & (Iowa)  & &  & &  & \multicolumn{2}{l}{D: legs/wings(PFA)}  \\
  &  & & & & &  & &  & \multicolumn{2}{l}{T : saliva(PFA)} \\
34 & WNV  & WNVLin\_CulPip\_BabMar\_8.2\_28\_A & lineage 1 & Culex pipiens pipiens  & 8,2 & 27  & 3,7,14,19  & 19 & I : abdomen (PFA)  & \citep{zakhia_experimental_2018} \\
  &  & & (France)  & (Liban)  & &  & &  & \multicolumn{2}{l}{D: legs(PFA)} \\
  &  & & & & &  & &  & \multicolumn{2}{l}{T : saliva(PFA)} \\
35 & WNV  & WNVLin\_albo\_StrBas\_6.7\_28\_A & lineage 1 & Aedes albopictus & 6,7 & 28  & 3,7,10,14,17,21,28 & 24 & I : abdomen (PFA)  & \citep{martinet_assessing_2023} \\
  &  & & (France)  & (France)  & &  & &  & \multicolumn{2}{l}{D: head(PFA)} \\
  &  & & & & &  & &  & \multicolumn{2}{l}{T : saliva(PFA)} \\
36 & WNV  & WNVNY9\_CulTar\_CalUSA\_7\_27\_A & NY99 , NY99-3557 & Culex tarsalis & 7 & 27  & 5,7,9,14  & 172 & I : abdomen (PFA)  & \citep{moudy_newly_2007}  \\
  &      &                                      & (West America)    & (USA)           &   &     &           &     & \multicolumn{2}{l}{D: legs(PFA)} \\
  &      &                                      &                   &                 &   &     &           &     & \multicolumn{2}{l}{T : saliva(PFA)} \\
37 & WNV  & WNVWN0\_CulTar\_CalUSA\_7\_27\_A & 02-1986 , 02-1956, Am  & Culex tarsalis & 7 & 27  & 5,7,9,14  & 185 & I : abdomen (PFA)  & \citep{moudy_newly_2007} \\
  &  & & (North America) & (USA)  & &  & &  & \multicolumn{2}{l}{D: legs(PFA)} \\
  &  & & & & &  & &  & \multicolumn{2}{l}{T : saliva(PFA)} \\ \bottomrule
\end{tabular}
}
\label{wnv_table}
\end{sidewaystable}

\begin{sidewaystable}[hbt!]
\centering
\caption{Summary of experimental data used to infer the three IVD model for Zika virus (ZIKV). ID:Infectious Dose (Log10 FFU/mL), FFA : Fluorescent Focus Assay, RT\-PCR:reverse transcription polymerase chain reaction Dpe: Day post exposition, ND:not defined }
\resizebox{\textwidth}{!}{%

\begin{tabular}{@{}lllllllllll@{}}
\toprule
\begin{tabular}[c]{@{}l@{}} Experimental conditions \\ \end{tabular} & Virus species & Experimental condition name & \begin{tabular}[c]{@{}l@{}}Virus strain \\ (origin)\end{tabular} & \begin{tabular}[c]{@{}l@{}}Mosquito genus \\ and species\\ (origin)\end{tabular} & ID & \begin{tabular}[c]{@{}l@{}}Ambiance conditions: \\ Temperature (°c)\\ Humidity (\%) \\ Light-dark cycle (hours)\end{tabular} & Dpe & \begin{tabular}[c]{@{}l@{}}Mosquito number \\ (mean by Dpe)\end{tabular} & \begin{tabular}[c]{@{}l@{}}Mosquitoes parts analyzed for \\ Infection (I) \\ Dissemination(D) \\ Transmission (T) \\ (method)\end{tabular} & Ref \\ \midrule
38 & ZIKV & FrG\_2016\_aeg\_Ncal\_7 & SA-2016-18246 & Ae.aegypti & 7 & 28°c & 6,9,14,21 & 29 & I:abdomen/thorax (PFA) & \citep{calvez_differential_2018} \\
 & & & (French Guiana) & \multicolumn{2}{l}{(New Caledonia)} & 80 & & & \multicolumn{2}{l}{D:head PFA)} \\
 & & & & & & 12h:12h & & & \multicolumn{2}{l}{T:saliva (PFA)} \\
39 & ZIKV & FrP\_aeg\_Tah\_6.8 & ZIKV strain PF13/251013-18 & Ae.aegypti & 7 (TCID50/mL)=6.8 & 27°c & 6,9,14,21 & 40 & I : abdomen/thorax (RT-PCR) & \citep{richard_vector_2016-1} \\
 & & & (French Polynesia) & \multicolumn{2}{l}{(Tahiti island, Toahotu)} & 80 & & & \multicolumn{2}{l}{D: legs(RT-PCR)} \\
 & & & & & & 12h:12h & & & \multicolumn{2}{l}{T : saliva(FFA)} \\
40 & ZIKV & Ncal\_2014\_aeg\_Ncal\_7 & NC-2014-843 & Ae.aegypti & 7 & 28°c & 6,9,14,21 & 30 & I:abdomen/thorax (PFA) & \citep{calvez_differential_2018} \\
 & & & (New Caledonia, 2014) & \multicolumn{2}{l}{(New Caledonia)} & 80 & & & \multicolumn{2}{l}{D:head PFA)} \\
 & & & & & & 12h:12h & & & \multicolumn{2}{l}{T:saliva (PFA)} \\
41 & ZIKV & Ncal\_2015\_aeg\_Ncal\_7 & NC-2015-2391 & Ae.aegypti & 7 & 28°c & 6,9,14,21 & 34 & I:abdomen/thorax (PFA) & \citep{calvez_differential_2018} \\
 & & & (New Caledonia, 2015) & \multicolumn{2}{l}{(New Caledonia)} & 80 & & & \multicolumn{2}{l}{D:head PFA)} \\
 & & & & & & 12h:12h & & & \multicolumn{2}{l}{T:saliva (PFA)} \\
42 & ZIKV & Ncal\_aeg\_FrP\_7 & NC-2014-5132 & Ae.aegypti & 7 & 28°c & 6,9,14,21 & 32 & I:abdomen/thorax (PFA) & \citep{calvez_zika_2018} \\
 & & & (New Caledonia) & \multicolumn{2}{l}{(French Polynesia)} & 80 & & & \multicolumn{2}{l}{D:head PFA)} \\
 & & & & & & 12h:12h & & & \multicolumn{2}{l}{T:saliva (PFA)} \\
43 & ZIKV & Ncal\_aeg\_Ncal\_7 & NC-2014-5132 & Ae.aegypti & 7 & 28°c & 6,9,14,21 & 26 & I:abdomen/thorax (PFA) & \citep{calvez_zika_2018} \\
 & & & (New Caledonia) & \multicolumn{2}{l}{(New Caledonia)} & 80 & & & \multicolumn{2}{l}{D:head PFA)} \\
 & & & & & & 12h:12h & & & \multicolumn{2}{l}{T:saliva (PFA)} \\
44 & ZIKV & Ncal\_aeg\_Sam\_7 & NC-2014-5132 & Ae.aegypti & 7 & 28°c & 6,9,14,21 & 39 & I:abdomen/thorax (PFA) & \citep{calvez_zika_2018} \\
 & & & (New Caledonia) & (Samoa) & & 80 & & & \multicolumn{2}{l}{D:head PFA)} \\
 & & & & & & 12h:12h & & & \multicolumn{2}{l}{T:saliva (PFA)} \\
45 & ZIKV & Ncal\_albo\_Rab\_7.2 & NC-2014-5132 & Ae.albopictus & 7.2 & 28 +/- 1°c & 3,7,14,21 & 28 & I: abdomen/thorax (PFA) & \citep{amraoui_potential_2019} \\
 & & & (New Caledonia) & \multicolumn{2}{l}{(Morocco, Rabat)} & 80 & & & \multicolumn{2}{l}{D:head (PFA)} \\
 & & & & & & 16h:8h & & & \multicolumn{2}{l}{T:saliva (PFA)} \\
46 & ZIKV & Ncal\_albo\_Tun\_7 & NC-2014-5132 & Ae.albopictus & 7 & 28 +/- 1°c & 7,10,14,21 & 23 & I:abdomen (PFA) & \citep{bohers_recently_2020} \\
 & & & (New Caledonia) & \multicolumn{2}{l}{(Tunisia,Car, Ami, La Mar)} & 80 & & & \multicolumn{2}{l}{D:thorax/head (PFA)} \\
 & & & & & & 16h:8h & & & \multicolumn{2}{l}{T:saliva (PFA)} \\
47 & ZIKV & Ncal\_pol\_FrP\_7 & NC-2014-5132 & Ae.polynesiensis & 7 & 28°c & 6,9,14,21 & 30 & I:abdomen/thorax (PFA) & \citep{calvez_zika_2018} \\
 & & & (New Caledonia) & \multicolumn{2}{l}{(French Polynesia)} & 80 & & & \multicolumn{2}{l}{D:head PFA)} \\
 & & & & & & 12h:12h & & & \multicolumn{2}{l}{T:saliva (PFA)} \\
48 & ZIKV & Ncal\_pol\_Wal\_7 & NC-2014-5132 & Ae.polynesiensis & 7 & 28°c & 6,9,14,21 & 38 & I:abdomen/thorax (PFA) & \citep{calvez_zika_2018} \\
 & & & (New Caledonia) & (Wallis) & & 80 & & & \multicolumn{2}{l}{D:head PFA)} \\
 & & & & & & 12h:12h & & & \multicolumn{2}{l}{T:saliva (PFA)} \\
49 & ZIKV & PueRic\_aeg\_PozRic\_7.2 & PRVABC59 & Ae.aegypti & 7.2 & 28°c & 2,4,6,8,10,12,14,16,18,20 & 30 & I:midgut(RT-PCR) & \citep{robison_comparison_2020} \\
 & & & (United States Puerto Rico) & \multicolumn{2}{l}{(Mexico, Poza Rica)} & 70-80 & & & \multicolumn{2}{l}{D:legs/wings (RT-PCR)} \\
 & & & & & & 12h:12h & & & \multicolumn{2}{l}{T:saliva (PFA)} \\
50 & ZIKV & Sen\_1991\_aeg\_Ncal\_7 & AF-1991-HD78788 & Ae.aegypti & 7 & 28°c & 6,9,14,21 & 28 & I:abdomen/thorax (PFA) & \citep{calvez_differential_2018} \\
 & & & (Senegal, 1991) & \multicolumn{2}{l}{(New Caledonia)} & 80 & & & \multicolumn{2}{l}{D:head PFA)} \\
 & & & & & & 12h:12h & & & \multicolumn{2}{l}{T:saliva (PFA)} \\
51 & ZIKV & Sen\_2002\_aeg\_Ncal\_7 & AF-2002-ArD 165 522 & Ae.aegypti & 7 & 28°c & 6,9,14,21 & 30 & I:abdomen/thorax (PFA) & \citep{calvez_differential_2018} \\
 & & & (Senegal, 2002) & \multicolumn{2}{l}{(New Caledonia)} & 80 & & & \multicolumn{2}{l}{D:head PFA)} \\
 & & & & & & 12h:12h & & & \multicolumn{2}{l}{T:saliva (PFA)} \\
52 & ZIKV & Ugan\_aeg\_Tow\_6.5 & MR 766 & Ae.aegypti & 6.7 +/- 0.2 TCID50=6.5 & 28°c & 5,7,10,14 & 24 & I:body (RT-PCR) & \citep{hall-mendelin_assessment_2016} \\
 & & & (Uganda) & \multicolumn{2}{l}{(Australia,Townsville)} & \multicolumn{2}{l}{high relative Humidity} & & \multicolumn{2}{l}{D:legs/wings (RT-PCR)} \\
 & & & & & & 12h:12h & & & \multicolumn{2}{l}{T:saliva (RT-PCR)} \\ \bottomrule
\end{tabular}
}
\label{zikv_table}
\end{sidewaystable}

\begin{table}[hbt!]
\caption{Pairwise comparisons of model RMSE across the 10 selected experimental conditions. For each experimental conditions, the table reports the RMSE difference between two models ($\Delta$RMSE $=$ RMSE$_{\mathrm{model\,A}}-$RMSE$_{\mathrm{model\,B}}$) with its 95\% credible interval (q025--q975), and the posterior probability that each model yields a lower RMSE. Positive $\Delta$RMSE values indicate a lower RMSE for model B.}
\label{pairwise_rmse}
\begin{tabular}{@{}llllll@{}}
\toprule
Experimental conditions & Model A & Model B & $\Delta$RMSE (q025--q975) & {Pr(RMSE$_A$ < RMSE$_B$)} & {Pr(RMSE$_B$ < RMSE$_A$)}\\ \midrule
1 & EIDT & EIDTP & 2,07 (2,03–2,11) & 0,00 & 1,00 \\
  & EIDT & EIDTPT & 1,56 (1,51–1,60) & 0,02 & 0,98 \\
  & EIDTP & EIDTPT & -0,51 (-0,57– -0,46) & 0,72 & 0,28 \\
3 & EIDT & EIDTP & 1,42 (1,36–1,47) & 0,06 & 0,94 \\
  & EIDT & EIDTPT & 1,47 (1,42–1,52) & 0,05 & 0,95 \\
  & EIDTP & EIDTPT & 0,05 (0,00–0,11) & 0,48 & 0,52 \\
4 & EIDT & EIDTP & 0,45 (0,42–0,49) & 0,28 & 0,72 \\
  & EIDT & EIDTPT & 0,44 (0,40–0,48) & 0,29 & 0,71 \\
  & EIDTP & EIDTPT & -0,01 (-0,03–0,01) & 0,52 & 0,48 \\
6 & EIDT & EIDTP & 0,77 (0,73–0,81) & 0,14 & 0,86 \\
  & EIDT & EIDTPT & 0,68 (0,64–0,72) & 0,16 & 0,84 \\
  & EIDTP & EIDTPT & -0,09 (-0,14– -0,04) & 0,54 & 0,46 \\
7 & EIDT & EIDTP & -0,04 (-0,07– -0,01) & 0,55 & 0,45 \\
  & EIDT & EIDTPT & 0,00 (-0,03–0,03) & 0,50 & 0,50 \\
  & EIDTP & EIDTPT & 0,04 (0,01–0,07) & 0,46 & 0,54 \\
8 & EIDT & EIDTP & -0,05 (-0,08– -0,02) & 0,55 & 0,45 \\
  & EIDT & EIDTPT & -0,09 (-0,11– -0,06) & 0,58 & 0,42 \\
  & EIDTP & EIDTPT & -0,04 (-0,07– -0,01) & 0,53 & 0,47 \\
9 & EIDT & EIDTP & 0,28 (0,25–0,31) & 0,31 & 0,69 \\
  & EIDT & EIDTPT & 0,30 (0,27–0,34) & 0,29 & 0,71 \\
  & EIDTP & EIDTPT & 0,02 (-0,01–0,06) & 0,47 & 0,53 \\
30 & EIDT & EIDTP & 0,07 (0,04–0,10) & 0,44 & 0,56 \\
  & EIDT & EIDTPT & 0,02 (-0,02–0,05) & 0,48 & 0,52 \\
  & EIDTP & EIDTPT & -0,06 (-0,09– -0,02) & 0,53 & 0,47 \\
34 & EIDT & EIDTP & -0,45 (-0,49– -0,40) & 0,74 & 0,26 \\
  & EIDT & EIDTPT & -0,97 (-1,02– -0,92) & 0,86 & 0,14 \\
  & EIDTP & EIDTPT & -0,52 (-0,58– -0,46) & 0,70 & 0,30 \\
49 & EIDT & EIDTP & -0,18 (-0,21– -0,15) & 0,62 & 0,38 \\
  & EIDT & EIDTPT & -0,12 (-0,15– -0,09) & 0,63 & 0,37 \\
  & EIDTP & EIDTPT & 0,06 (0,03–0,09) & 0,50 & 0,50 \\ \bottomrule
\end{tabular}
\end{table}

\FloatBarrier

\begin{longtable}{@{}llllll@{}}
\caption{Pairwise comparisons of model RMSE across the 42 selected experimental conditions. For each experimental conditions, the table reports the RMSE difference between two models ($\Delta$RMSE $=$ RMSE$_{\mathrm{model\,A}}-$RMSE$_{\mathrm{model\,B}}$) with its 95\% credible interval (q025--q975), and the posterior probability that each model yields a lower RMSE. Positive $\Delta$RMSE values indicate a lower RMSE for model B.}
\label{pairwise_rmse_42}\\
\toprule
Experimental conditions & Model A & Model B & $\Delta$RMSE (q025--q975) & {Pr(RMSE$_A$ < RMSE$_B$)} & {Pr(RMSE$_B$ < RMSE$_A$)}\\
\midrule
\endfirsthead

\caption[]{Pairwise comparisons of model RMSE across the 42 selected experimental conditions(continued).}\\
\toprule
Experimental conditions & Model A & Model B & $\Delta$RMSE (q025--q975) & {Pr(RMSE$_A$ < RMSE$_B$)} & {Pr(RMSE$_B$ < RMSE$_A$)}\\
\midrule
\endhead

\midrule
\multicolumn{6}{r}{Continued on next page}\\
\endfoot

\bottomrule
\endlastfoot

2        & EIDT    & EIDTP   & -0,14 (-0,19 -- -0,10)  & 0,58422 & 0,41578 \\
         & EIDT    & EIDTPT  & -0,10 (-0,14 -- -0,05)  & 0,55512 & 0,44488 \\
         & EIDTP   & EIDTPT  & 0,05 (0,01 -- 0,09)     & 0,47124 & 0,52876 \\
5        & EIDT    & EIDTP   & -0,03 (-0,06 -- -0,01)  & 0,53312 & 0,46688 \\
         & EIDT    & EIDTPT  & -0,05 (-0,07 -- -0,03)  & 0,55508 & 0,44492 \\
         & EIDTP   & EIDTPT  & -0,01 (-0,04 -- 0,01)   & 0,5267  & 0,4733  \\
10       & EIDT    & EIDTP   & -0,61 (-0,68 -- -0,55)  & 0,73436 & 0,26564 \\
         & EIDT    & EIDTPT  & -0,90 (-0,96 -- -0,83)  & 0,80664 & 0,19336 \\
         & EIDTP   & EIDTPT  & -0,29 (-0,36 -- -0,22)  & 0,59506 & 0,40494 \\
11       & EIDT    & EIDTP   & -0,05 (-0,08 -- -0,03)  & 0,5272  & 0,4728  \\
         & EIDT    & EIDTPT  & -0,08 (-0,11 -- -0,05)  & 0,52884 & 0,47116 \\
         & EIDTP   & EIDTPT  & -0,02 (-0,06 -- 0,01)   & 0,50104 & 0,49896 \\
12       & EIDT    & EIDTP   & 0,00 (-0,07 -- 0,07)    & 0,51    & 0,49    \\
         & EIDT    & EIDTPT  & -0,08 (-0,16 -- -0,01)  & 0,53334 & 0,46666 \\
         & EIDTP   & EIDTPT  & -0,08 (-0,16 -- -0,01)  & 0,52328 & 0,47672 \\
13       & EIDT    & EIDTP   & 0,00 (-0,03 -- 0,03)    & 0,49674 & 0,50326 \\
         & EIDT    & EIDTPT  & 0,00 (-0,03 -- 0,03)    & 0,50906 & 0,49094 \\
         & EIDTP   & EIDTPT  & 0,00 (-0,03 -- 0,03)    & 0,50828 & 0,49172 \\
14       & EIDT    & EIDTP   & 0,05 (0,01 -- 0,09)     & 0,46712 & 0,53288 \\
         & EIDT    & EIDTPT  & 0,08 (0,04 -- 0,12)     & 0,45242 & 0,54758 \\
         & EIDTP   & EIDTPT  & 0,03 (-0,01 -- 0,07)    & 0,4865  & 0,5135  \\
15       & EIDT    & EIDTP   & -0,34 (-0,38 -- -0,29)  & 0,66074 & 0,33926 \\
         & EIDT    & EIDTPT  & -0,17 (-0,22 -- -0,12)  & 0,58506 & 0,41494 \\
         & EIDTP   & EIDTPT  & 0,17 (0,12 -- 0,22)     & 0,42574 & 0,57426 \\
16       & EIDT    & EIDTP   & 0,12 (0,08 -- 0,16)     & 0,42718 & 0,57282 \\
         & EIDT    & EIDTPT  & -0,21 (-0,26 -- -0,16)  & 0,60902 & 0,39098 \\
         & EIDTP   & EIDTPT  & -0,33 (-0,38 -- -0,29)  & 0,67756 & 0,32244 \\
17       & EIDT    & EIDTP   & 0,13 (0,08 -- 0,19)     & 0,4461  & 0,5539  \\
         & EIDT    & EIDTPT  & 0,13 (0,07 -- 0,19)     & 0,44606 & 0,55394 \\
         & EIDTP   & EIDTPT  & 0,00 (-0,05 -- 0,06)    & 0,49822 & 0,50178 \\
18       & EIDT    & EIDTP   & 0,06 (0,01 -- 0,10)     & 0,46584 & 0,53416 \\
         & EIDT    & EIDTPT  & 0,04 (-0,01 -- 0,08)    & 0,485   & 0,515   \\
         & EIDTP   & EIDTPT  & -0,02 (-0,06 -- 0,02)   & 0,5165  & 0,4835  \\
19       & EIDT    & EIDTP   & -0,12 (-0,19 -- -0,04)  & 0,53252 & 0,46748 \\
         & EIDT    & EIDTPT  & -0,08 (-0,15 -- -0,01)  & 0,51978 & 0,48022 \\
         & EIDTP   & EIDTPT  & 0,04 (-0,04 -- 0,11)    & 0,48752 & 0,51248 \\
20       & EIDT    & EIDTP   & 0,28 (0,00 -- 0,59)     & 0,46764 & 0,53236 \\
         & EIDT    & EIDTPT  & 0,16 (-0,13 -- 0,46)    & 0,48114 & 0,51886 \\
         & EIDTP   & EIDTPT  & -0,12 (-0,45 -- 0,17)   & 0,51326 & 0,48674 \\
21       & EIDT    & EIDTP   & 0,17 (-0,03 -- 0,37)    & 0,484   & 0,516   \\
         & EIDT    & EIDTPT  & -0,30 (-0,52 -- -0,07)  & 0,522   & 0,478   \\
         & EIDTP   & EIDTPT  & -0,46 (-0,68 -- -0,25)  & 0,53962 & 0,46038 \\
22       & EIDT    & EIDTP   & -0,01 (-0,21 -- 0,20)   & 0,49644 & 0,50356 \\
         & EIDT    & EIDTPT  & -0,16 (-0,37 -- 0,04)   & 0,51984 & 0,48016 \\
         & EIDTP   & EIDTPT  & -0,15 (-0,35 -- 0,06)   & 0,52092 & 0,47908 \\
23       & EIDT    & EIDTP   & -0,06 (-0,19 -- 0,06)   & 0,51616 & 0,48384 \\
         & EIDT    & EIDTPT  & -0,10 (-0,23 -- 0,03)   & 0,52098 & 0,47902 \\
         & EIDTP   & EIDTPT  & -0,04 (-0,16 -- 0,08)   & 0,50806 & 0,49194 \\
24       & EIDT    & EIDTP   & 0,21 (0,09 -- 0,33)     & 0,45786 & 0,54214 \\
         & EIDT    & EIDTPT  & 0,06 (-0,07 -- 0,19)    & 0,48558 & 0,51442 \\
         & EIDTP   & EIDTPT  & -0,15 (-0,29 -- -0,02)  & 0,53116 & 0,46884 \\
25       & EIDT    & EIDTP   & 0,00 (-0,17 -- 0,17)    & 0,50194 & 0,49806 \\
         & EIDT    & EIDTPT  & 0,04 (-0,13 -- 0,20)    & 0,49848 & 0,50152 \\
         & EIDTP   & EIDTPT  & 0,04 (-0,13 -- 0,20)    & 0,49498 & 0,50502 \\
26       & EIDT    & EIDTP   & -0,16 (-0,22 -- -0,10)  & 0,55216 & 0,44784 \\
         & EIDT    & EIDTPT  & -0,03 (-0,08 -- 0,03)   & 0,508   & 0,492   \\
         & EIDTP   & EIDTPT  & 0,13 (0,07 -- 0,20)     & 0,45724 & 0,54276 \\
27       & EIDT    & EIDTP   & 0,24 (0,11 -- 0,39)     & 0,45822 & 0,54178 \\
         & EIDT    & EIDTPT  & 0,22 (0,09 -- 0,37)     & 0,46018 & 0,53982 \\
         & EIDTP   & EIDTPT  & -0,02 (-0,15 -- 0,12)   & 0,50304 & 0,49696 \\
28       & EIDT    & EIDTP   & -0,19 (-0,29 -- -0,08)  & 0,54194 & 0,45806 \\
         & EIDT    & EIDTPT  & -0,41 (-0,51 -- -0,29)  & 0,57306 & 0,42694 \\
         & EIDTP   & EIDTPT  & -0,22 (-0,33 -- -0,10)  & 0,5344  & 0,4656  \\
29       & EIDT    & EIDTP   & -0,08 (-0,12 -- -0,05)  & 0,56614 & 0,43386 \\
         & EIDT    & EIDTPT  & -0,11 (-0,14 -- -0,08)  & 0,57734 & 0,42266 \\
         & EIDTP   & EIDTPT  & -0,03 (-0,06 -- 0,00)   & 0,51066 & 0,48934 \\
31       & EIDT    & EIDTP   & -0,19 (-0,22 -- -0,16)  & 0,64894 & 0,35106 \\
         & EIDT    & EIDTPT  & -0,26 (-0,29 -- -0,23)  & 0,6995  & 0,3005  \\
         & EIDTP   & EIDTPT  & -0,07 (-0,11 -- -0,04)  & 0,56402 & 0,43598 \\
32       & EIDT    & EIDTP   & -0,08 (-0,12 -- -0,05)  & 0,55594 & 0,44406 \\
         & EIDT    & EIDTPT  & -0,07 (-0,11 -- -0,03)  & 0,53156 & 0,46844 \\
         & EIDTP   & EIDTPT  & 0,01 (-0,03 -- 0,05)    & 0,48356 & 0,51644 \\
33       & EIDT    & EIDTP   & 0,08 (0,03 -- 0,14)     & 0,46524 & 0,53476 \\
         & EIDT    & EIDTPT  & 0,02 (-0,04 -- 0,07)    & 0,5028  & 0,4972  \\
         & EIDTP   & EIDTPT  & -0,07 (-0,12 -- -0,01)  & 0,53644 & 0,46356 \\
35       & EIDT    & EIDTP   & -0,05 (-0,07 -- -0,03)  & 0,57572 & 0,42428 \\
         & EIDT    & EIDTPT  & -0,04 (-0,05 -- -0,02)  & 0,5538  & 0,4462  \\
         & EIDTP   & EIDTPT  & 0,01 (0,00 -- 0,03)     & 0,47654 & 0,52346 \\
36       & EIDT    & EIDTP   & 0,13 (0,05 -- 0,21)     & 0,4553  & 0,5447  \\
         & EIDT    & EIDTPT  & 0,13 (0,05 -- 0,21)     & 0,4605  & 0,5395  \\
         & EIDTP   & EIDTPT  & 0,00 (-0,08 -- 0,07)    & 0,50748 & 0,49252 \\
37       & EIDT    & EIDTP   & -1,08 (-1,24 -- -0,93)  & 0,62244 & 0,37756 \\
         & EIDT    & EIDTPT  & -0,18 (-0,28 -- -0,08)  & 0,54734 & 0,45266 \\
         & EIDTP   & EIDTPT  & 0,90 (0,74 -- 1,05)     & 0,41288 & 0,58712 \\
38       & EIDT    & EIDTP   & 0,21 (0,12 -- 0,31)     & 0,44826 & 0,55174 \\
         & EIDT    & EIDTPT  & 0,11 (0,01 -- 0,21)     & 0,47196 & 0,52804 \\
         & EIDTP   & EIDTPT  & -0,10 (-0,20 -- -0,01)  & 0,52394 & 0,47606 \\
39       & EIDT    & EIDTP   & 0,06 (0,01 -- 0,11)     & 0,47352 & 0,52648 \\
         & EIDT    & EIDTPT  & 0,01 (-0,05 -- 0,06)    & 0,48256 & 0,51744 \\
         & EIDTP   & EIDTPT  & -0,05 (-0,10 -- 0,00)   & 0,5081  & 0,4919  \\
40       & EIDT    & EIDTP   & 0,00 (-0,09 -- 0,09)    & 0,49912 & 0,50088 \\
         & EIDT    & EIDTPT  & 0,06 (-0,02 -- 0,15)    & 0,47858 & 0,52142 \\
         & EIDTP   & EIDTPT  & 0,07 (-0,02 -- 0,15)    & 0,48096 & 0,51904 \\
41       & EIDT    & EIDTP   & 0,06 (-0,02 -- 0,14)    & 0,48762 & 0,51238 \\
         & EIDT    & EIDTPT  & 0,16 (0,08 -- 0,25)     & 0,45818 & 0,54182 \\
         & EIDTP   & EIDTPT  & 0,11 (0,03 -- 0,19)     & 0,4756  & 0,5244  \\
42       & EIDT    & EIDTP   & 0,02 (-0,05 -- 0,08)    & 0,4926  & 0,5074  \\
         & EIDT    & EIDTPT  & -0,20 (-0,26 -- -0,13)  & 0,58056 & 0,41944 \\
         & EIDTP   & EIDTPT  & -0,21 (-0,27 -- -0,15)  & 0,58774 & 0,41226 \\
43       & EIDT    & EIDTP   & -0,83 (-0,95 -- -0,69)  & 0,6192  & 0,3808  \\
         & EIDT    & EIDTPT  & -0,68 (-0,81 -- -0,56)  & 0,5992  & 0,4008  \\
         & EIDTP   & EIDTPT  & 0,14 (-0,03 -- 0,30)    & 0,47894 & 0,52106 \\
44       & EIDT    & EIDTP   & 0,07 (-0,02 -- 0,15)    & 0,48738 & 0,51262 \\
         & EIDT    & EIDTPT  & -0,24 (-0,34 -- -0,15)  & 0,5592  & 0,4408  \\
         & EIDTP   & EIDTPT  & -0,31 (-0,40 -- -0,21)  & 0,5747  & 0,4253  \\
45       & EIDT    & EIDTP   & 0,19 (0,12 -- 0,27)     & 0,44124 & 0,55876 \\
         & EIDT    & EIDTPT  & 0,16 (0,08 -- 0,24)     & 0,44964 & 0,55036 \\
         & EIDTP   & EIDTPT  & -0,03 (-0,11 -- 0,04)   & 0,50924 & 0,49076 \\
46       & EIDT    & EIDTP   & 0,00 (-0,04 -- 0,04)    & 0,49318 & 0,50682 \\
         & EIDT    & EIDTPT  & 0,03 (-0,01 -- 0,07)    & 0,4751  & 0,5249  \\
         & EIDTP   & EIDTPT  & 0,03 (-0,01 -- 0,08)    & 0,47978 & 0,52022 \\
47       & EIDT    & EIDTP   & -0,07 (-0,14 -- 0,00)   & 0,52802 & 0,47198 \\
         & EIDT    & EIDTPT  & -0,10 (-0,17 -- -0,03)  & 0,52964 & 0,47036 \\
         & EIDTP   & EIDTPT  & -0,03 (-0,10 -- 0,04)   & 0,50348 & 0,49652 \\
48       & EIDT    & EIDTP   & -0,02 (-0,21 -- 0,16)   & 0,49726 & 0,50274 \\
         & EIDT    & EIDTPT  & 0,01 (-0,18 -- 0,21)    & 0,49496 & 0,50504 \\
         & EIDTP   & EIDTPT  & 0,03 (-0,17 -- 0,23)    & 0,49892 & 0,50108 \\
50       & EIDT    & EIDTP   & -0,03 (-0,07 -- 0,01)   & 0,52722 & 0,47278 \\
         & EIDT    & EIDTPT  & -0,10 (-0,13 -- -0,06)  & 0,57434 & 0,42566 \\
         & EIDTP   & EIDTPT  & -0,07 (-0,10 -- -0,03)  & 0,54952 & 0,45048 \\
51       & EIDT    & EIDTP   & -0,15 (-0,24 -- -0,05)  & 0,53306 & 0,46694 \\
         & EIDT    & EIDTPT  & -0,20 (-0,30 -- -0,10)  & 0,54808 & 0,45192 \\
         & EIDTP   & EIDTPT  & -0,05 (-0,15 -- 0,05)   & 0,51492 & 0,48508 \\
52       & EIDT    & EIDTP   & -0,10 (-0,13 -- -0,07)  & 0,55582 & 0,44418 \\
         & EIDT    & EIDTPT  & -0,06 (-0,09 -- -0,03)  & 0,5476  & 0,4524  \\
         & EIDTP   & EIDTPT  & 0,04 (0,00 -- 0,07)     & 0,48886 & 0,51114 \\
\end{longtable}

\begin{figure}[hbt!]

\centering
\includegraphics[width=\textwidth ,keepaspectratio]{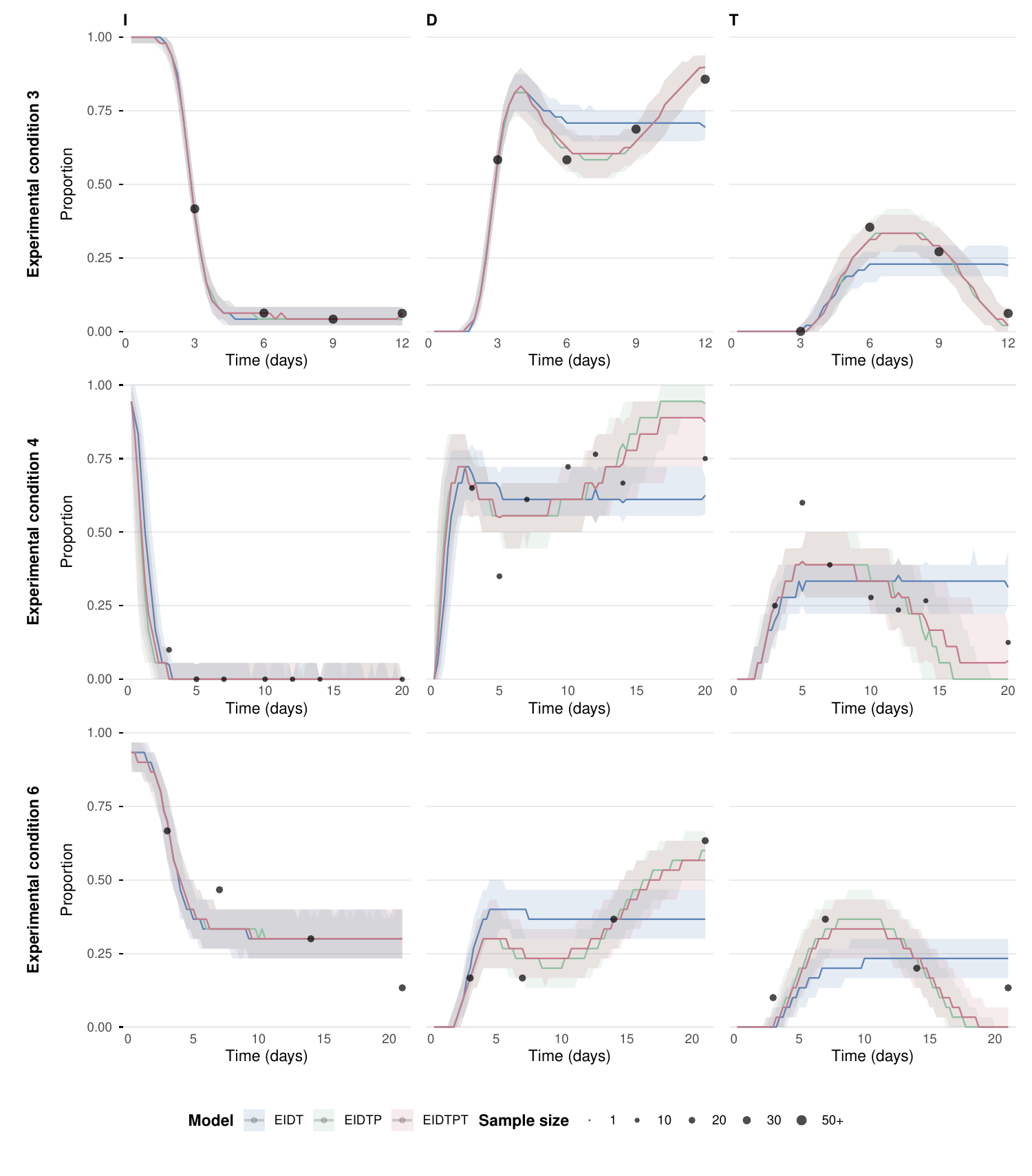}
\caption{\textbf{Comparison of visual fit between the three models (EIDT): permanent transmission, (EIDTP): definitive interruption of transmission, and (EIDTPT): temporary interruption of transmission.}
Observed and simulated dynamics of the proportions of infected ($I$), disseminated ($D$), and transmitter ($T$) mosquitoes over time for the three models for experimental conditions 3,4,6 .
Points represent observed mosquitoes proportions at each sampling time, with point size proportional to the sample size. The solid line shows the median, and the shaded envelope indicates the uncertainty band spanning the 25th to the 75th percentile.}
\label{visual_fit_supp}

\end{figure}

\begin{figure}[hbt!]
\centering
\includegraphics[width=\textwidth]{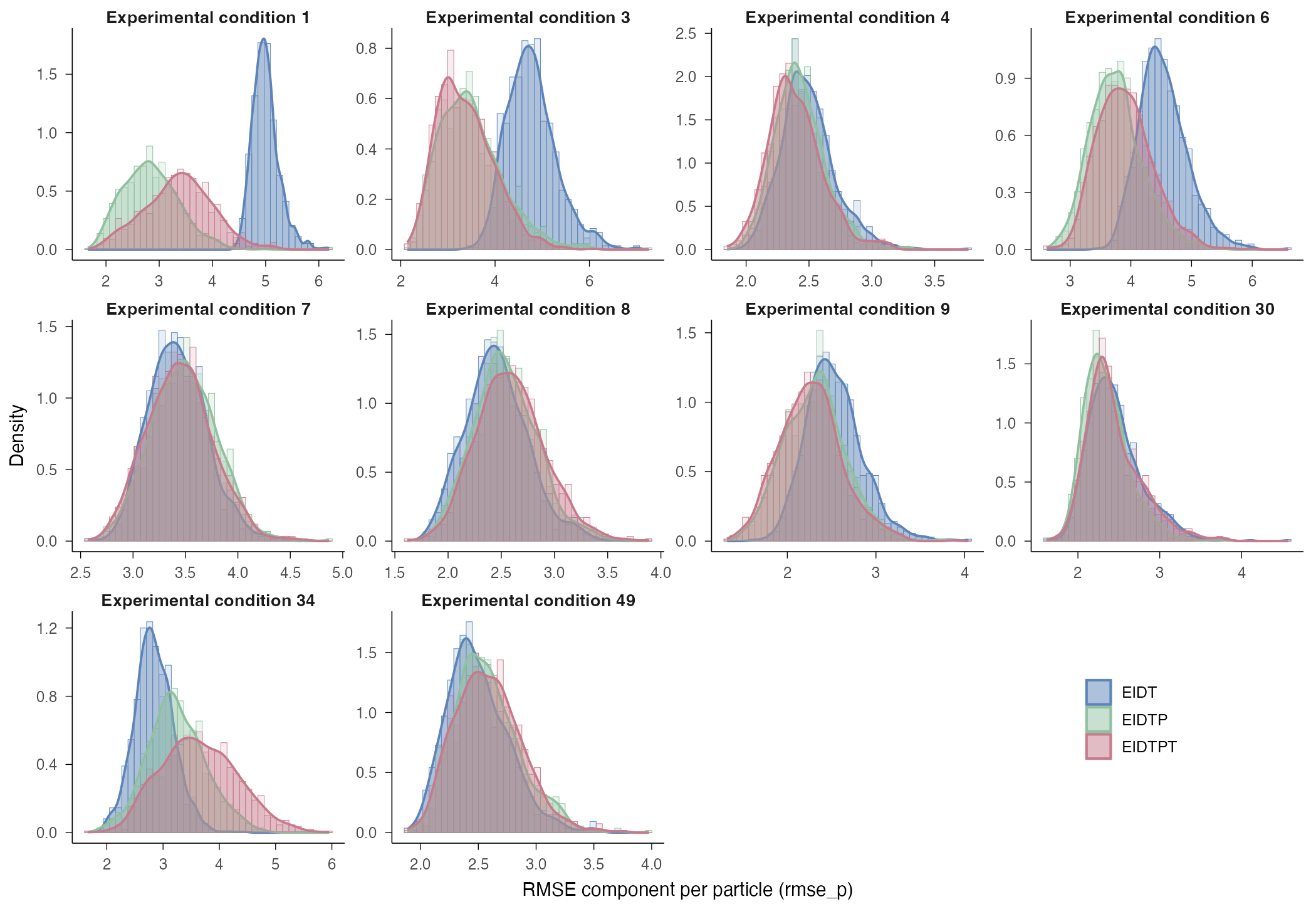}
\caption{\textbf{Weighted distributions of RMSE components across selected experimental conditions. }
Each panel corresponds to one experimental condition and shows the distribution of particle-specific RMSE component values for the three competing models (EIDT, EIDTP, EIDTPT). Histograms and kernel density curves are weighted by normalized posterior particle weights, so that each model-specific distribution integrates to one within an experimental condition. Colors identify models consistently across panels. This representation allows visual comparison of the location, spread, and overlap of posterior RMSE component distributions among models for each selected experimental condition.}
\label{rmse_distributions}
\end{figure}

\begin{figure}[hbt!]
\centering
\includegraphics[width=\textwidth]{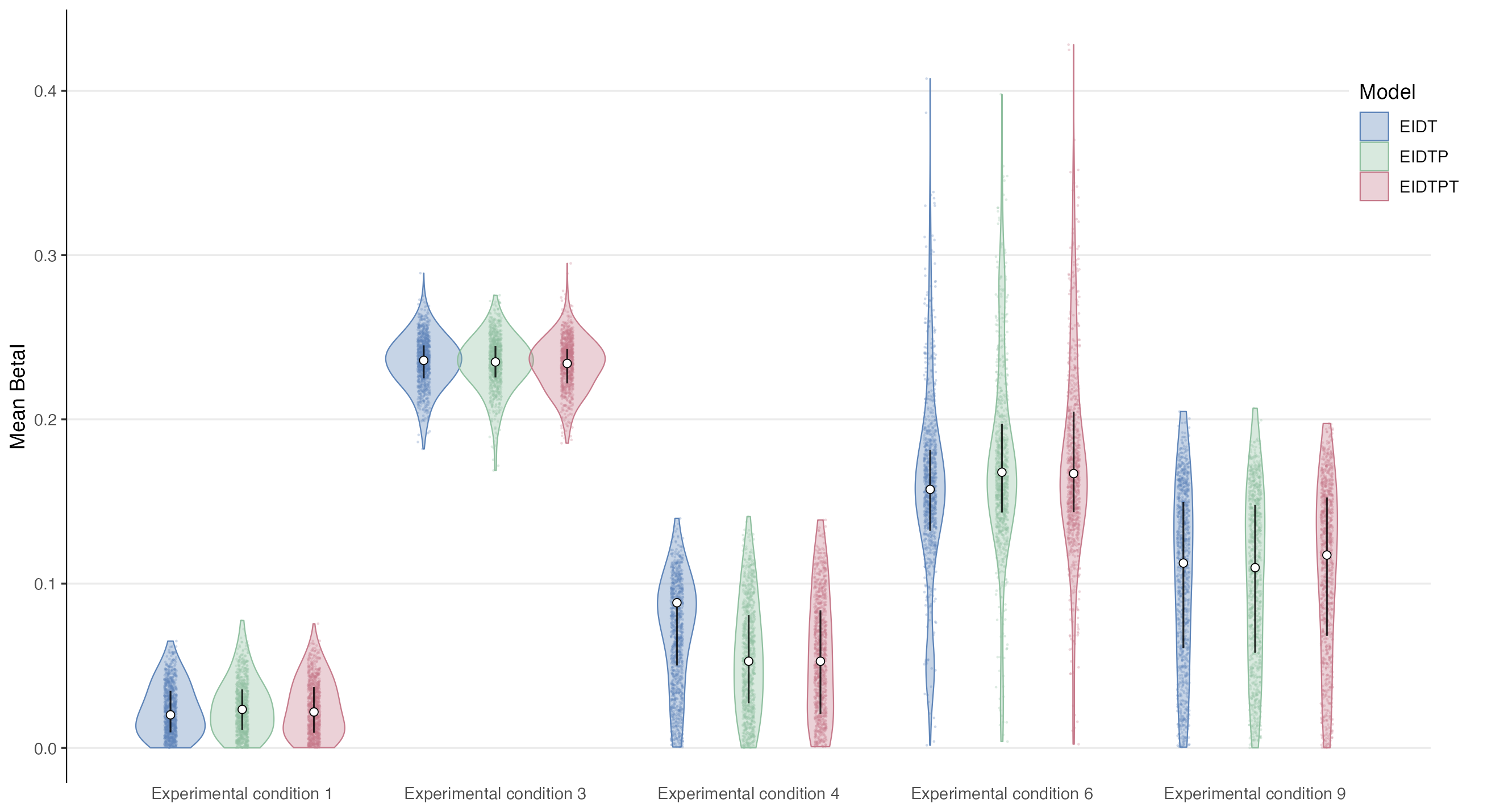}
\caption{\textbf{Posterior distributions of the mean duration in infected state across experimental conditions 1, 3, 4, 6, 9 and model structures (EIDT: permanent transmission; EIDTP: definitive interruption of transmission; EIDTPT: temporary interruption of transmission)}. Violin plots represent distribution of mean duration in infected state for each model and experimental conditions, with white dots indicating the median and vertical bars the interquartile range (q25–q75).}
\label{violin_plot_betaI_mean}

\end{figure}

\begin{figure}[hbt!]
\centering
\includegraphics[width=\textwidth]{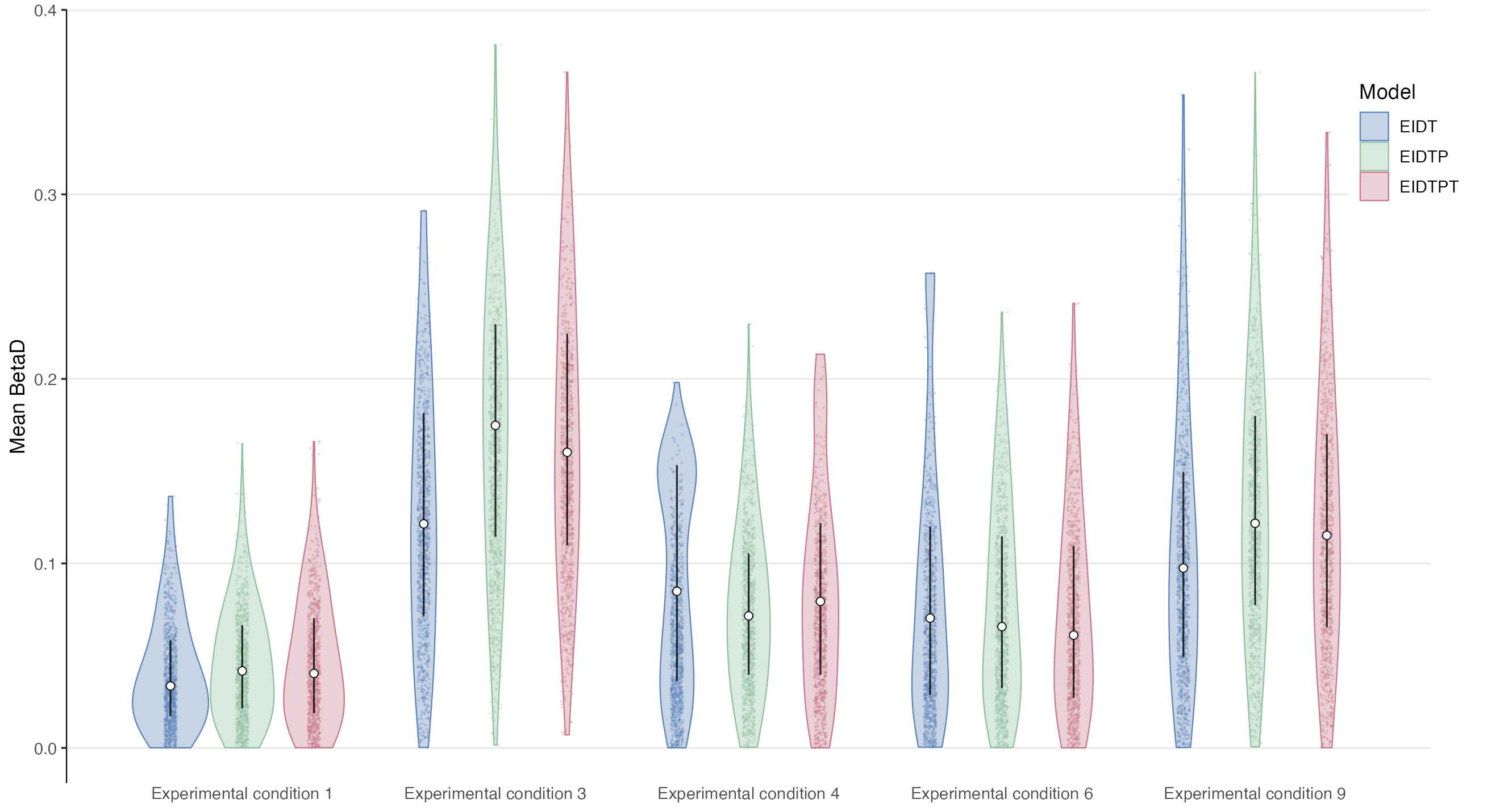}
\caption{\textbf{Posterior distributions of the mean duration in disseminated state across experimental conditions  1, 3, 4, 6, 9 and model structures (EIDT: permanent transmission; EIDTP: definitive interruption of transmission; EIDTPT: temporary interruption of transmission)}. Violin plots represent distribution of mean duration in disseminated state for each model and experimental condition, with white dots indicating the median and vertical bars the interquartile range (q25–q75).}
\label{violin_plot_betaD_mean}

\end{figure}

\begin{figure}[hbt!]
\centering
\includegraphics[width=0.9\textwidth]{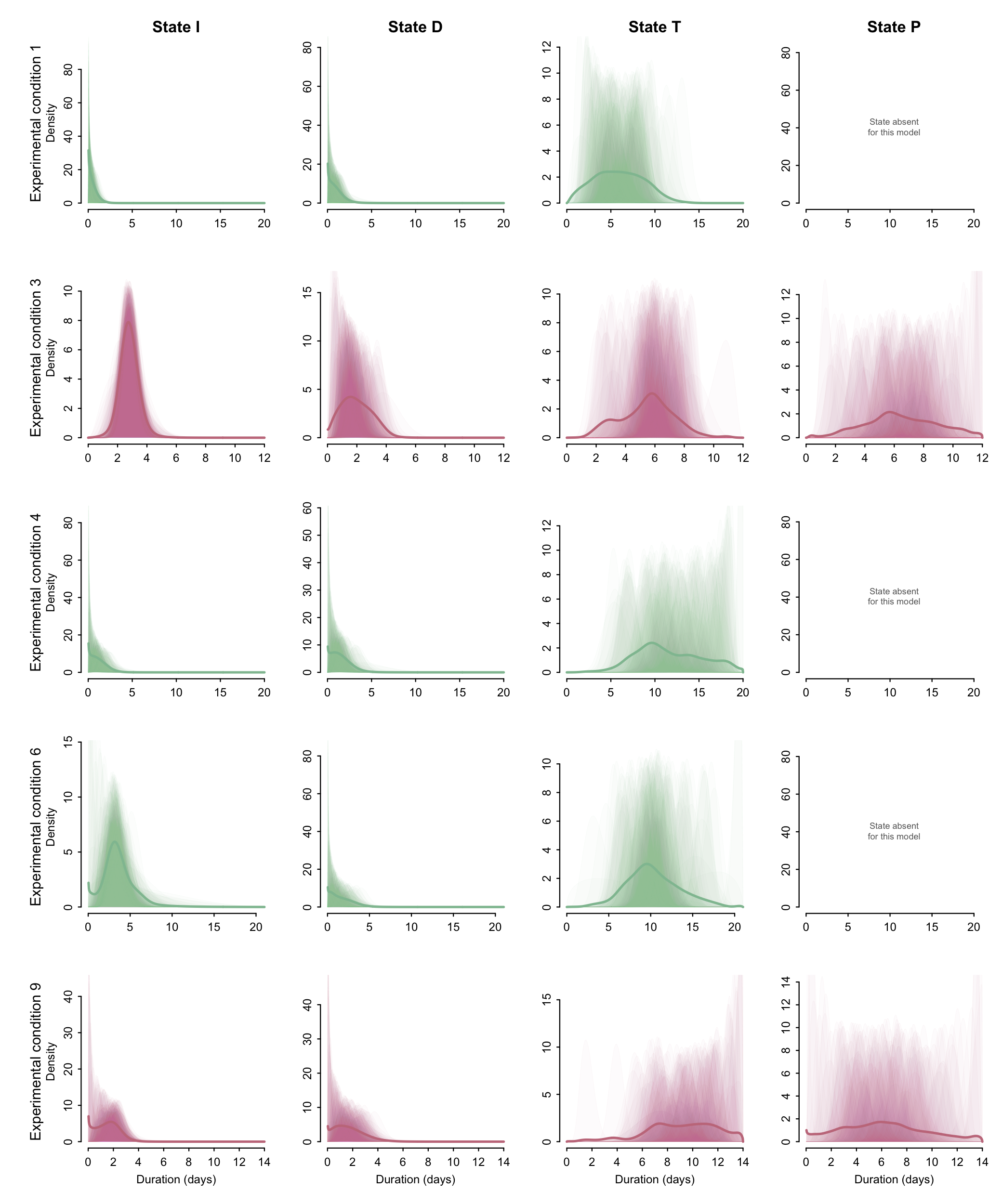}
\caption{\textbf{Distributions of state residence times for the selected experimental conditions and their best-supported models. }
Each row corresponds to one selected experimental condition and each column to a state associated with a residence-time distribution when present in the selected model ($I$, $D$, $T$, $P$). Within each panel, thin semi-transparent curves represent posterior residence-time densities for individual accepted particles, whereas the thick curve represents the weighted posterior mean density. Colors indicate the model selected for each experimental condition. The x-axis is expressed in days, obtained by scaling the corresponding beta-distribution support by the observed experimental condition duration. Panels labelled “State absent for this model” indicate that the corresponding state-specific residence-time distribution is not defined in the selected model. For readability, only a random subset of posterior particle densities may be displayed, while the mean density is always computed from the full posterior sample.}
\label{duration_in_IDTP_detailed}
\end{figure}

\begin{figure}[hbt!]
\centering
\includegraphics[width=0.7\textwidth]{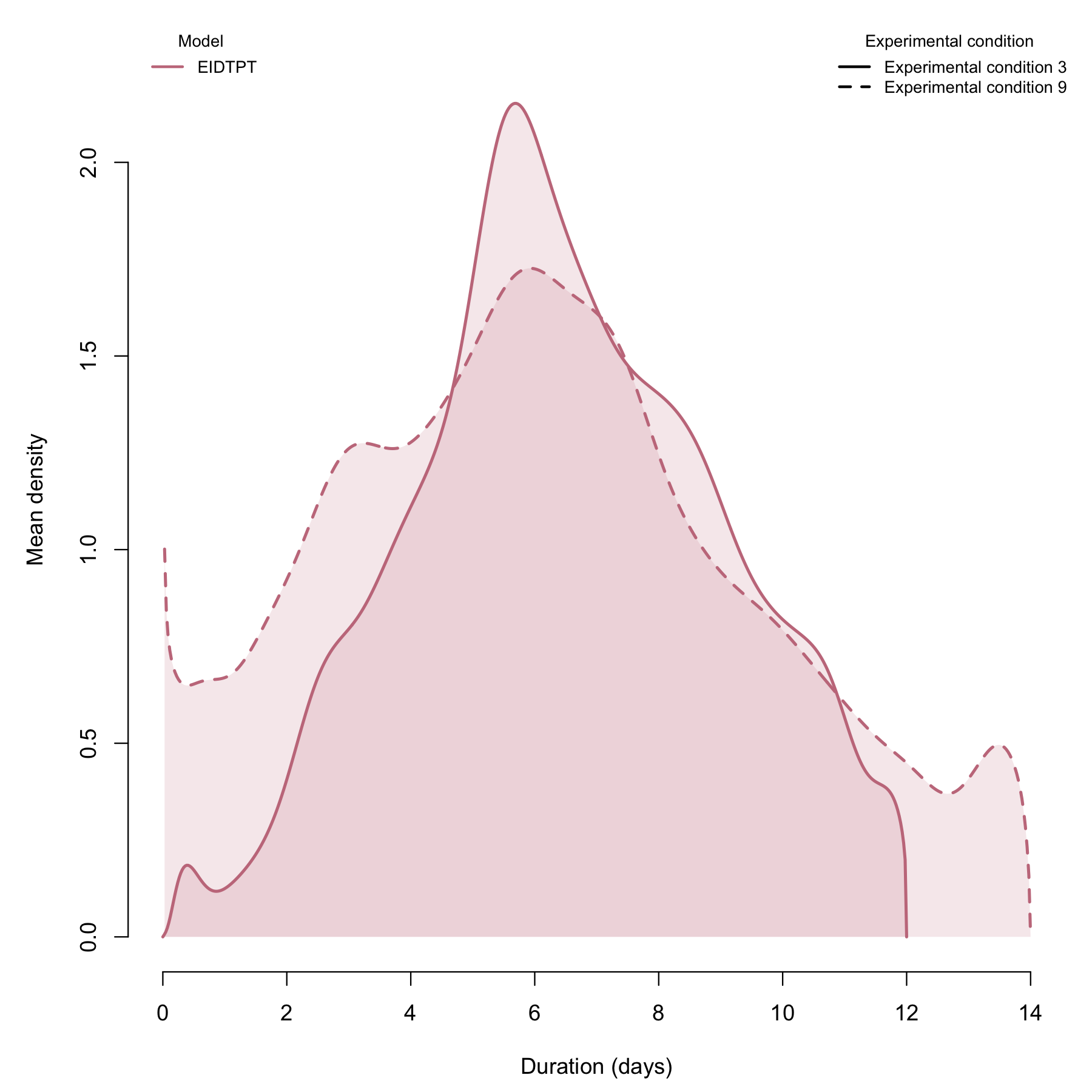}
\caption{\textbf{Distribution of non-transmitting-state duration in experimental conditions best explained by the intermittent interruption model. }
Distributions of the duration spent in the $P$ state (temporary interruption of transmission) for experimental conditions 3 and 9, estimated under models allowing temporary transmission interruption (EIDTPT). Densities represent the distribution of the mean duration in state $P$ (in days).}
\label{mean_duration_P_detailed}

\end{figure}

\end{document}